\documentclass[aps,pra,floatfix,preprint,onecolumn,superscriptaddress]{revtex4-2}
\usepackage{graphicx}
\usepackage{amsmath, amsfonts, amssymb, bm}
\usepackage{amstext}
\usepackage{mathrsfs}
\usepackage{hyperref}
\makeatletter
\newcommand{\gig}{\bBigg@{3}}
\makeatother

\usepackage{xcolor}

\begin{document}
\title{Infrared behavior of the photon yield in nonlinear Compton scattering}
\author{Antonino Di Piazza}
\email{a.dipiazza@rochester.edu}
\affiliation{Department of Physics and Astronomy, University of Rochester, Rochester, New York 14627, USA}
\affiliation{Laboratory for Laser Energetics, University of Rochester, Rochester, New York 14623, USA}
\author{Giulio Audagnotto}
\email{giulio.audagnotto@unito.it}
\affiliation{Department of Physics and Astronomy, University of Rochester, Rochester, New York 14627, USA}
\affiliation{Dipartimento di Fisica, Universit\`a di Torino \& INFN, Sezione di Torino, \\
Via Pietro Giuria 1, I-10125 Turin, Italy}

\begin{abstract}
Nonlinear Compton scattering is the process of emission of a single photon by a charge driven by an intense laser field. Here, we study the infrared behavior of the photon yield emitted via nonlinear Compton scattering. We first consider the idealized case of an electron in the presence of a plane wave and derive an analytical expression of the total yield in the form of a double integral over the laser phase. As it is known, the total yield is finite in the experimentally common case of a plane-wave laser field without a dc component whereas it diverges logarithmically in the complementary case of so-called unipolar fields. The divergence is found here to correspond to the longer-and-longer formation lengths of emitted photons with lower-and-lower frequency. Interestingly, we also find that the corrections to the Volkov states stemming from the fact that the field is unipolar cancel out in the computation of the probability of nonlinear Compton scattering, which is in agreement with the fact that in the classical limit the expression of the photon yield is independent of whether the plane wave is unipolar or not. Then, we pass to the more realistic case of an electron in a tightly-focused laser beam by assuming that the electron is ultrarelativistic. In this case we determine analytically the classical angular distribution of the yield of photons with an energy larger than a fixed value $\hbar\omega_m$ as a double integral over the electron's trajectory. Indeed, in the case of a focused laser beam the electron does typically undergo a net acceleration and it is therefore appropriate to fix a lower bound to avoid the infrared divergence. After obtaining also the corresponding quantum expression of the angular distribution of the photon yield within the quasiclassical approximation, we determine analytically the leading-order quantum correction, which scales as $\hbar\omega_m/\varepsilon$, where $\varepsilon$ is the initial electron energy. Considerations on the experimental feasibility of the found results in the latter situations are also presented.
\end{abstract}

\maketitle
\thispagestyle{empty}

\clearpage
\section{Introduction}
An accelerated charge emits electromagnetic radiation \cite{Jackson_b_1975,Landau_b_2_1975}. Classically, the emission is theoretically allowed for arbitrary angular frequencies $\omega$, whereas quantum mechanically an upper limit is set by the fact that a photon with angular frequency $\omega$ carries an energy $\hbar\omega$ (quantum recoil) \cite{Landau_b_4_1982}. Here, we are specifically interested in the low-energy regime $\hbar\omega \rightarrow 0$, in which the radiation is ``soft''. The emission of radiation with arbitrary low frequencies depends on the global acceleration that the charge undergoes and not on the local properties of its motion. In other words, the radiation energy spectrum from a point-like charge with initial four-momentum $p^\alpha$ and final four-momentum $p'^\alpha$ after the interaction with an external agent, is different from zero in the soft limit $\omega \rightarrow 0$ if and only if $p'^\alpha \neq p^\alpha$  \cite{Jackson_b_1975}. When this limit is not zero, the quantum mechanical photon number spectrum, or the classical energy spectrum divided by $\hbar\omega$, diverges logarithmically \cite{Jackson_b_1975,Landau_b_2_1975,Peskin_b_1995}.
These considerations can be readily proven classically, recalling that the spectrum of radiation with four-momentum $k^\alpha = \omega n^\alpha$ by a point-like charge depends on the square modulus of the four-current (units with $\epsilon_0=\hbar=c=1$ are used here and below, such that $\alpha=e^2/4\pi\approx 1/137$ is the fine-structure constant) \cite{Jackson_b_1975, Landau_b_2_1975}
\begin{equation}
\begin{split}
J^\alpha(k) = i e\int d \tau \frac{d}{d \tau } \left(
\frac{\pi^\alpha (\tau)}{(k \pi (\tau))}
\right) e ^{i (k x(\tau))}   ,
\end{split}
\end{equation}
where $x^\alpha(\tau)$ and $\pi^\alpha(\tau)$ are the particle's four-position and four-momentum, respectively, as functions of the charge proper time $\tau$ and where $(ab)$ indicates the four-dimensional scalar product between two generic four-vectors $a^{\mu}$ and $b^{\mu}$ (the metric tensor $\eta_{\mu\nu}=\text{diag}(+1,-1,-1,-1)$ is assumed). For the sake of definiteness, we have implicitly assumed the charge to be an electron with charge $e<0$ and mass $m$.

It is manifest that for the four-current to have a zero-frequency Fourier component, the electron has to undergo a net (or global) acceleration. Indeed, in the low-energy limit, the familiar expression \cite{Jackson_b_1975, Peskin_b_1995} 
\begin{equation}
\label{IR limit of the current}
\begin{split}
J^\alpha(k) \overset{\text{IR}}{=} i e
\left(
\frac{p'^\alpha}{(kp')}
- 
\frac{p^\alpha}{(kp)}
\right),
\end{split}
\end{equation}
is obtained, showing that $J^\alpha(k) \overset{\text{IR}}{\sim} 1/\omega $. This in turn implies that in the infrared limit the energy spectrum tends to a constant value $d \mathcal{E} \propto |J(k)|^2 d^3 \bm{k} \overset{\text{IR}}{\sim} C d \omega d \Omega $.

The above classical considerations only rely on the given trajectory of the electron. We now consider an electron driven by an external electromagnetic field. In the case of a general field configuration, the electron will undergo a net acceleration. Ideal plane waves represent an exception: when the plane wave does not feature a dc component and when the interaction of the electron with its own field is neglected, the electron does not undergo a net acceleration. We recall that the self interaction of a particle with its own electromagnetic field goes under the name of radiation reaction. This phenomenon has been widely studied in the literature for several decades (apart from the original works in Refs. \cite{Dirac_1938, Abraham_b_1905}, we refer the reader to the reviews \cite{Di_Piazza_2012,Burton_2014,Gonoskov_2022,Fedotov_2023}) but only recently measured \cite{Cole_2018, Poder_2018, Wistisen_2018, Los2026}. 

The classical emission from a charge in a strong laser field is known as nonlinear Thomson scattering and a detailed analysis of its soft limit including radiation reaction in a plane wave has been presented in Ref. \cite{Di_Piazza_2018_b}. Such a configuration is theoretically interesting for its mathematical simplicity and experimentally relevant as an idealized model for laser-particle interactions. Here, we first study the soft limit of such a process, including the possibility of the plane wave featuring a dc component (unipolar fields). Moreover, we analyze the emission in a more general field configuration and its connection with the semiclassical limit of nonlinear Compton scattering, i.e., the emission of a single photon in a plane wave including quantum effects \cite{Di_Piazza_2012,Fedotov_2023}.

Although the infrared divergence of the photon number has no physical meaning from a classical point of view, in a quantum perspective it underlines an intrinsic interpretational problem. In fact, the number of photons is an observable quantity and, as such, one would expect it to be unambiguously finite (see below). Besides this, perturbation theory in quantum field theory brings additional issues in the infrared sector and infrared divergences have left substantial traces in the literature \cite{Bloch_1937, Yennie_1961,Weinberg:1965nx, Agarwal_2021}. Indeed, it has been noticed since the early developments of QED that bremsstrahlung amplitudes lead to infrared divergent probabilities. As a typical example, one can consider the scattering of an electron by a heavy charged particle. Leaving aside the heavy scatterer whose dynamics can be ignored, one finds a well-known result: the amplitude for the emission of a soft photon with four-momentum $k^\mu$ and polarization four-vector $e^{\mu}_{l}(k)$ by the electron factorizes by a term proportional to $(J(k)e^*_{l}(k))$ (see Eq. \eqref{IR limit of the current}) \cite{Bloch_1937, Peskin_b_1995}:
\begin{equation}
\label{soft factorization amplitude}
\begin{split}
M_{\gamma} \overset{\text{IR}}{=} e \left[
\frac{ (p' e^*_{l}(k) ) }{ (k  p')}
- 
\frac{ ( p  e^*_{l}(k) )}{(k  p)}
\right]M_{0},
\end{split}
\end{equation}
where $M_{\gamma}$ ($M_0$) is the amplitude for the process with (without) the emission of the real soft photon. This amplitude can be squared and integrated to find the corresponding cross section.
In the process of calculating the total cross section, however, a logarithmic divergence arises, which needs to be regularized. This is a clear difference as compared to the classical framework, and it is also the source of a problem, which is not interpretational but structural. If a theory is plagued with ultraviolet divergences, in fact, one can eliminate them with a renormalization procedure \cite{Peskin_b_1995,Itzykson_b_1980}. Clearly, this cannot be carried out for the infrared divergences, which are large-scale (low-energy) limits of the theory, and as such they have to reproduce the classical results. The solution of such issues in QED comes from the Bloch-Nordsieck theorem \cite{Bloch_1937}, which basically shows how real bremsstrahlung infrared divergences cancel out with virtual corrections, such that the observable cross sections are independent of the regularization parameters. Following this procedure, one obtains a finite cross section, which, however, depends on the minimum photon energy measurable by the detector (detector resolution) and it diverges logarithmically for a detector with infinite resolution \cite{Peskin_b_1995,Itzykson_b_1980}. Indeed, massless quanta in an infinite volume can have arbitrarily small energy (large wavelength) and the integrated probability of emitting such particles does grow logarithmically. Nevertheless, to measure arbitrarily small energies, i.e., large wavelengths, one would need an arbitrarily large experimental apparatus. Thus, the measurement of zero-frequency radiation is as unphysical as requiring an infinitely large photon detector. This is a quite unique feature as the probability of an elementary process does depend on the characteristics of the measuring device.

We now consider QED in a strong electromagnetic background. Strong-field QED (SF-QED) describes the properties of a quantum system interacting with a strong electromagnetic field, which is treated as a classical background field. By ``strong'' one means that the background field measured in the reference system of the charged particles like the electrons is comparable with the corresponding Schwinger critical field $F_{cr} = m^2 /|e| = 1.3 \times 10^{16} \, \textrm{V/cm}$ \cite{Di_Piazza_2012,Gonoskov_2022,Fedotov_2023}. This requirement is expressed by means of the quantum nonlinearity parameter $\chi= \sqrt{-(p_\mu F^{\mu\nu})^2}\, /mF_{cr}$ via the condition $\chi \gtrsim 1 $. Here, $F^{\mu\nu}$ is the constant Maxwell tensor characterizing the peak amplitude of the background electromagnetic field.

In the following, we will study the infrared limit of both nonlinear Thomson and Compton scattering. The infrared behavior of the spectrum in nonlinear Compton scattering has also been studied in Refs. \cite{Dinu_2012, Ilderton_2013_b,King_2021,Edwards_2021,Krachkov_2024}. We will see how the presence of a dc component can be treated within the plane-wave model, as previously investigated in Refs.
\cite{Dinu_2012, Ilderton_2013_b}. We will show how the production of soft photons and the soft divergences can be interpreted in terms of the longer-and-longer formation lengths of photons with larger-and-larger wavelengths. Additionally, the more realistic setup of an electron interacting with a tightly-focused laser beam will be analyzed both analytically and numerically. In the SF-QED case this is carried out within the quasiclassical method, where the electron energy is assumed to be the largest dynamical energy in the problem \cite{Di_Piazza_2014,Di_Piazza_2015,Di_Piazza_2021}.

The paper is organized as follows: In Sec. \ref{Sec. Integrated photon yield via nonlinear Compton scattering} the angularly integrated spectrum of nonlinear Compton scattering in a plane wave is analyzed, with a particular focus on its infrared limit. Also, the connection between soft radiation and large formation length is established and the generalizations of the previous considerations including radiation-reaction effects is discussed. Section \ref{Sec. Classical angular distribution of radiation} is devoted to the study of the angular distribution of photons emitted with a frequency larger than a fixed value. 
Section \ref{Sec. Nonlinear Compton scattering probability in an unipolar plane-wave field} deals with the issue of defining asymptotic in- and out-states in a unipolar plane-wave field and with non-linear Compton scattering in such fields. Finally, Sec. \ref{Sec. Emitted photon yield in an arbitrary field within the quasiclassical approximation} extends the results of Sec. \ref{Sec. Classical angular distribution of radiation} to the case of nonlinear Compton scattering by an ultrarelativistic electron in an arbitrary background field and the first quantum correction to the infrared limit of the spectrum is calculated within the quasiclassical approximation. The main conclusions of the paper are presented in Sec. \ref{Conclusions}. An appendix contains technical details on the calculation of the probability of nonlinear Compton scattering in a tightly-focused laser field within the quasiclassical approximation.

\section{Integrated photon yield via nonlinear Compton scattering in a plane wave}
\label{Sec. Integrated photon yield via nonlinear Compton scattering}
In a plane-wave background, the differential probability $dP/d^3\bm{k}$ of photon emission via nonlinear Compton scattering per unit of photon momentum $\bm{k}$ is well known and has been studied in several publications (see the reviews \cite{Mitter_1975,Ritus_1985,Ehlotzky_2009,Reiss_2009,Di_Piazza_2012,Gonoskov_2022,Fedotov_2023} and the more recent papers \cite{Torgrimsson_2024,Zhao_2025,Di_Piazza_2026}). We refer here to the results presented in Ref. \cite{Di_Piazza_2018}, as we will use the same notation.

We consider an electron with initial four-momentum $p^{\mu}=(\varepsilon,\bm{p})$, colliding with a plane wave whose propagation direction is denoted by $\bm{n}_L$ ($\bm{n}_L^2=1$). The plane wave is described by the four-vector potential $A^{\mu}(\phi)=(0,\bm{A}_{\perp}(\phi))$, where $\phi=(nx)=t-\bm{n}_L\cdot\bm{x}$ (i.e., $n_L^{\mu}=(1,\bm{n}_L)$), and where $\bm{n}_L\cdot\bm{A}_{\perp}(\phi)=0$ and $\lim_{\phi\to-\infty}\bm{A}_{\perp}(\phi)=\bm{0}$  (i.e., the four-potential is chosen in the Lorent gauge $\partial_{\mu}A^{\mu}(\phi)=0$). We assume that the function $\bm{A}_{\perp}(\phi)$ is ``physically'' well behaved in the sense that it is continuous with its derivative everywhere \footnote{It is important to point out that the assumption on the continuity of $\bm{A}_{\perp}(\phi)$ is necessary even from a strictly mathematical point of view in order to carry out the change of variable from the time $t$ to the light-cone time $\phi$ in the integrals.}. For the moment, we do not make any assumption about the limit $\lim_{\phi\to+\infty}\bm{A}_{\perp}(\phi)=\bm{A}_{\perp}(\infty)$ because the value of $\bm{A}_{\perp}(\infty)$ distinguishes whether the plane wave is unipolar ($\bm{A}_{\perp}(\infty)\neq \bm{0}$) or not ($\bm{A}_{\perp}(\infty)= \bm{0}$). Since the plane wave depends only on the variable $\phi$, it is convenient to introduce the light-cone coordinates $T=(t+\bm{n}_L\cdot\bm{x})/2$, $\bm{x}_{\perp}=\bm{x}-(\bm{n}_L\cdot\bm{x})\bm{n}_L$, together with $\phi=t-\bm{n}_L\cdot\bm{x}$, as well as the light-cone components $v_+=(v^0+\bm{n}_L\cdot\bm{v})/2$, $\bm{v}_{\perp}=\bm{v}-(\bm{n}_L\cdot\bm{v})\bm{n}_L$, and $v_-=v^0-\bm{n}_L\cdot\bm{v}$ of an arbitrary four-vector $v^{\mu}=(v^0,\bm{v})$. Assuming that the emitted photon (outgoing electron) is characterized by a four-momentum $k^{\mu}=(\omega,\bm{k})$ ($p^{\prime\mu}=(\varepsilon',\bm{p}')$), the leading-order emission probability $dP/d^3\bm{k}$ averaged (summed) over all initial (final) discrete quantum numbers can be computed starting from the corresponding $S$-matrix element within the Furry picture \cite{Landau_b_4_1982}. By referring the reader to Ref. \cite{Di_Piazza_2018} for more details, we report here the final result in the form
\begin{equation}
\label{dP_d^3k}
\begin{split}
\frac{dP}{d^3\bm{k}}
=&-\frac{1}{4\pi^2}\frac{\alpha m^2}{p_-p'_-\omega}\int d\phi d\phi'\,e^{-i\frac{k_-m^2}{2p_-p'_-}\int_{\phi}^{\phi'}d\tilde{\phi}\,\left\{1+\left[\frac{\bm{p}_{\perp}}{m}-\frac{p_-}{k_-}\frac{\bm{k}_{\perp}}{m}-\bm{\xi}_{\perp}(\tilde{\phi})\right]^2\right\}}\\
&\quad\times\left\{1+\frac{1}{4}\frac{p_-^2+p_-^{\prime\,2}}{p_-p'_-}[\pmb{\xi}_{\perp}(\phi)-\pmb{\xi}_{\perp}(\phi')]^2\right\},
\end{split}
\end{equation}
where $p'_-=p_--k_-$ and $\pmb{\xi}_{\perp}(\phi)=e\bm{A}_{\perp}(\phi)/m$. By passing from the variable $k_{\parallel}=\bm{k}\cdot\bm{n}_L$ to the variable $k_-=\omega-k_{\parallel}=\sqrt{\bm{k}_{\perp}^2+k_{\parallel}^2}-k_{\parallel}$, the resulting integral over the transverse photon momenta $\bm{k}_{\perp}$ is Gaussian and can be taken analytically:
\begin{equation}
\begin{split}
\frac{dP}{dk_-}=&-i\frac{\alpha}{2\pi}\frac{1}{p_-}\frac{\xi_0}{\chi_0}\int\frac{d\varphi d\varphi'}{\varphi-\varphi'+i0}\bigg\{1+\frac{p_-^2+p^{\prime\,2}_-}{4p_-p'_-}[\pmb{\xi}_{\perp}(\varphi)-\pmb{\xi}_{\perp}(\varphi')]^2\bigg\}\\
&\times\exp\left\langle i\frac{1}{2}\frac{k_-}{p'_-}\frac{\xi_0}{\chi_0}\left\{\varphi-\varphi'+\int_{\varphi'}^{\varphi}d\tilde{\varphi}\,\pmb{\xi}^2_{\perp}(\tilde{\varphi})-\frac{1}{\varphi-\varphi'}\left[\int_{\varphi'}^{\varphi}d\tilde{\varphi}\,\pmb{\xi}_{\perp}(\tilde{\varphi})\right]^2\right\}\right\rangle,
\end{split}
\end{equation}
where the classical and the quantum nonlinearity parameters $\xi_0=|e|E_0/m\omega_0$ and $\chi_0=(p_-/m)E_0/F_{cr}$, with $E_0$ and $\omega_0$ being the amplitude of the electric field and the central angular frequency of the plane wave, respectively, have been introduced together with the laser phase $\varphi=\omega_0\phi$ ($\varphi'=\omega_0\phi'$). Also, the prescription $\varphi-\varphi'+i0$ results from requiring that the integral over the transverse photon momenta is absolutely convergent. This prescription also ensures that the emission probability vanishes if the external field vanishes. Finally, by passing to the variables $\varphi_+=(\varphi+\varphi')/2$ and $\varphi_-=\varphi-\varphi'$, we obtain
\begin{equation}
\begin{split}
\frac{dP}{dk_-}&=-i\frac{\alpha}{2\pi}\frac{1}{p_-}\frac{\xi_0}{\chi_0}\int\frac{d\varphi_+d\varphi_-}{\varphi_-+i0}\left\{1+\frac{p_-^2+p_-^{\prime\,2}}{4p_-p'_-}\left[\pmb{\xi}_{\perp}\left(\varphi_+-\frac{\varphi_-}{2}\right)-\pmb{\xi}_{\perp}\left(\varphi_++\frac{\varphi_-}{2}\right)\right]^2\right\}\\
&\quad\times\exp\left\langle\frac{i}{2}\frac{k_-}{p'_-}\frac{\xi_0}{\chi_0}\Bigg\{\varphi_-+\int_{-\varphi_-/2}^{\varphi_-/2}d\tilde{\varphi}\,\pmb{\xi}^2_{\perp}(\varphi_++\tilde{\varphi})-\frac{1}{\varphi_-}\bigg[\int_{-\varphi_-/2}^{\varphi_-/2}d\tilde{\varphi}\,\pmb{\xi}_{\perp}(\varphi_++\tilde{\varphi})\bigg]^2\Bigg\}\right\rangle.
\end{split}
\end{equation}

Now, we would like to take also the integral over $k_-$ but we should keep in mind that the result will only have a formal validity in the case of a unipolar plane wave because the corresponding total probability is (logarithmically) divergent. This is an example of soft or infrared divergences, which correspond to the fact that in a collision phenomenon an electron can potentially emit an infinite number of photons with lower-and-lower energy, such that the total emitted energy still remains finite \cite{Jackson_b_1975,Jauch_b_1976,Peskin_b_1995}. This can be recognized already at the classical level, if one computes the photon-emission yield by dividing the emission energy spectrum by the ``photon energy'' $\omega$ \cite{Jackson_b_1975}. Indeed, as we have already mentioned in the introduction, one easily recognizes that if the final momentum of the electron differs from the initial one, the differential number of photons linearly diverges in the limit $\omega\to 0$ such that the total photon yield diverges logarithmically. By looking at the exact four-momentum of an electron in a plane wave, one also sees that this occurs if and only if (in the present case) $\bm{A}_{\perp}(\infty)\neq \bm{A}_{\perp}(-\infty)=\bm{0}$ (see, in particular, Refs. \cite{Dinu_2012,Ilderton_2013_b,Krachkov_2024} for a detailed study of the infrared behavior in nonlinear Compton scattering in a plane wave).

By keeping in mind that the resulting expression is finite only if $\bm{A}_{\perp}(\infty)=\bm{0}$, the integral can be taken by passing to the variable $u=k_-/p'_-=k_-/(p_--k_-)$. After explicitly subtracting the (vanishing) probability at zero field, the total probability can be written as
\begin{equation}
\label{P}
\begin{split}
P&=\frac{\alpha}{2\pi}\frac{\xi_0}{\chi_0}\text{Im}\int\frac{d\varphi_+d\varphi_-}{\varphi_-}\int_0^{\infty}\frac{du}{1+u}\left\{\frac{e^{iu\Psi}-e^{iu\Psi_0}}{1+u}+\frac{1}{4}\left[1+\frac{1}{(1+u)^2}\right]\right.\\
&\quad\left.\times\left[\pmb{\xi}_{\perp}\left(\varphi_+-\frac{\varphi_-}{2}\right)-\pmb{\xi}_{\perp}\left(\varphi_++\frac{\varphi_-}{2}\right)\right]^2e^{iu\Psi}\right\},
\end{split}
\end{equation}
where we have removed the now unnecessary prescription at $\varphi_-=0$ and where
\begin{align}
\Psi&=\frac{1}{2}\frac{\xi_0}{\chi_0}\varphi_-\Bigg\{1+\frac{1}{\varphi_-}\int_{-\varphi_-/2}^{\varphi_-/2}d\tilde{\varphi}\,\pmb{\xi}^2_{\perp}(\varphi_++\tilde{\varphi})-\bigg[\frac{1}{\varphi_-}\int_{-\varphi_-/2}^{\varphi_-/2}d\tilde{\varphi}\,\pmb{\xi}_{\perp}(\varphi_++\tilde{\varphi})\bigg]^2\Bigg\},\\
\Psi_0&=\frac{1}{2}\frac{\xi_0}{\chi_0}\varphi_-.
\end{align}
The integrals in $u$ can be expressed in terms of the incomplete Gamma function $\Gamma(0,z)$ \cite{NIST_b_2010}, with $z$ being a complex number, because the following identities hold:
\begin{align}
I_1&=\text{Im}\int_0^{\infty}du\,\frac{e^{iu\Psi}-e^{iu\Psi_0}}{(1+u)^2}=\Psi\text{Re}\, e^{-i\Psi}\Gamma(0,-i\Psi)-\Psi_0 \text{Re}\, e^{-i\Psi_0}\Gamma(0,-i\Psi_0),\\
I_2&=\text{Im}\int_0^{\infty}du\,\frac{e^{iu\Psi}}{1+u}=\text{Im}\,e^{-i\Psi}\Gamma(0,-i\Psi),\\
I_3&=\text{Im}\int_0^{\infty}du\,\frac{e^{iu\Psi}}{(1+u)^3}=\frac{\Psi}{2}-\frac{\Psi^2}{2}\text{Im}\,e^{-i\Psi}\Gamma(0,-i\Psi).
\end{align}
The expression of the total probability in Eq. (\ref{P}), with the integrals $I_1$, $I_2$, and $I_3$ expressed in terms of the incomplete Gamma function, is valid for an arbitrary (non-unipolar, in the sense specified above) plane wave and various limiting expressions can be derived. 

For example, the corresponding classical expression is obtained in the limit $\chi_0\to 0$. By expanding the incomplete Gamma functions \cite{NIST_b_2010} in the asymptotic limit of large second argument, the resulting ``classical'' probability $P_c$ or photon yield is
\begin{equation}
\label{P_c}
\begin{split}
P_c&=\frac{\alpha}{\pi}\int\frac{d\varphi_+d\varphi_-}{\varphi_-^2}\left\{\frac{1}{2}\left[\pmb{\xi}_{\perp}\left(\varphi_+-\frac{\varphi_-}{2}\right)-\pmb{\xi}_{\perp}\left(\varphi_++\frac{\varphi_-}{2}\right)\right]^2\right.\\
&\quad\left.-\frac{1}{\varphi_-}\int_{-\varphi_-/2}^{\varphi_-/2}d\tilde{\varphi}\,\pmb{\xi}^2_{\perp}(\varphi_++\tilde{\varphi})+\bigg[\frac{1}{\varphi_-}\int_{-\varphi_-/2}^{\varphi_-/2}d\tilde{\varphi}\,\pmb{\xi}_{\perp}(\varphi_++\tilde{\varphi})\bigg]^2\right\}\\
&\quad\times\left\{1+\frac{1}{\varphi_-}\int_{-\varphi_-/2}^{\varphi_-/2}d\tilde{\varphi}\,\pmb{\xi}^2_{\perp}(\varphi_++\tilde{\varphi})-\bigg[\frac{1}{\varphi_-}\int_{-\varphi_-/2}^{\varphi_-/2}d\tilde{\varphi}\,\pmb{\xi}_{\perp}(\varphi_++\tilde{\varphi})\bigg]^2\right\}^{-1}.
\end{split}
\end{equation}
The probability $P_c$ can be equivalently calculated starting from the classical differential energy spectrum \cite{Jackson_b_1975}
\begin{equation}
\label{dE_dk}
\frac{d\mathcal{E}_c}{d^3\bm{k}}=\frac{\alpha}{4\pi^2}\left|\int dt\,\bm{n}\times[\bm{n}\times\bm{v}(t)]e^{i\omega\int_0^td\tilde{t}[1-\bm{n}\cdot\bm{v}(\tilde{t})]}\right|^2,
\end{equation}
emitted by an electron in the presence of the plane wave under consideration and dividing it by the photon energy $\omega$ before integrating it over $d^3\bm{k}$. Here, $\bm{n}=\bm{k}/\omega$ and $\bm{v}(t)$ is the electron velocity in the plane wave.

Analogously, one can compute the limiting expression for $\chi_0\gg 1$. In this case it is convenient to first exploit the symmetry of the integrand under the transformation $\varphi_-\to-\varphi_-$ to write the integral over $\varphi_-$ from zero to infinity. Again, by employing the asymptotic expression of the incomplete Gamma function for small second argument, one obtains
\begin{align}
\label{P_Large_chi}
P&=\frac{\alpha}{8}\frac{\xi_0}{\chi_0}\int d\varphi_+\int_0^{\infty}\frac{d\varphi_-}{\varphi_-}\left[\pmb{\xi}_{\perp}\left(\varphi_+-\frac{\varphi_-}{2}\right)-\pmb{\xi}_{\perp}\left(\varphi_++\frac{\varphi_-}{2}\right)\right]^2, && \text{for} && \chi_0\gg 1,
\end{align}
which is inversely proportional to $\chi_0$, or equivalently to the initial electron energy.

\begin{figure}
\begin{center}
\includegraphics[width=0.9\textwidth]{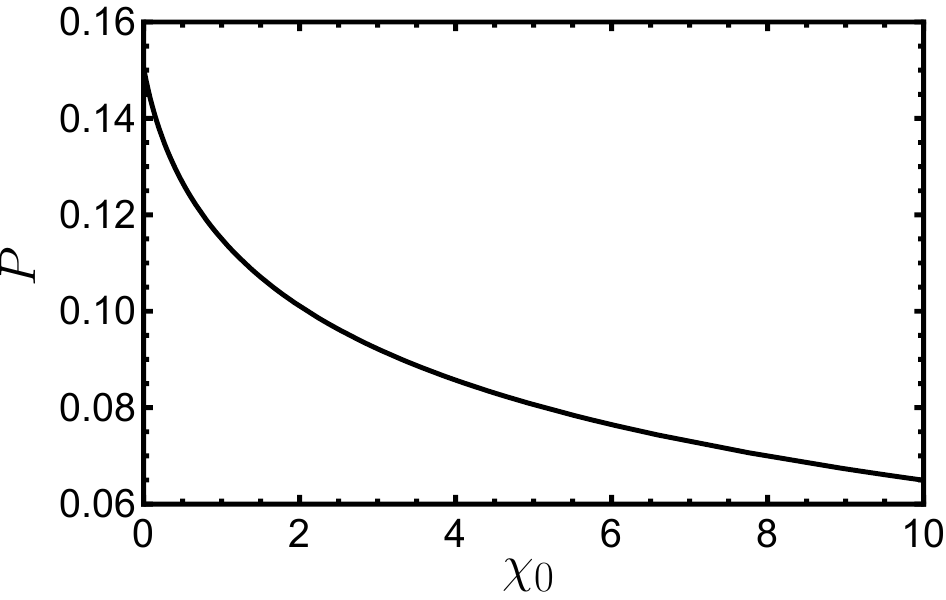}
\end{center}
\caption{Total photon yield $P$ as a function of $\chi_0$ for numerical parameters given in the text.}
\label{P_chi}
\end{figure}

In Fig. \ref{P_chi}, we plot the total yield $P$ as a function of $\chi_0$ in the case of a non-unipolar, linearly polarized plane wave with phase-dependent dimensionless amplitude given by $\xi(\varphi)=\xi_0\sinh(\varphi)/\cosh^2(\varphi)$, corresponding to a one-cycle pulse. We set $\xi_0=10$ and the minimum plotted value of $\chi_0$ is $0.01$. It can be seen that the value of the probability at $\chi_0=0.01$ differs from the value computed via $P_c$, equal to $0.15055$, by less than $1\,\%$. On the other hand, the asymptotic at large $\chi_0$ given in Eq. (\ref{P_Large_chi}) would provide a good approximation ($\sim 10\,\%$) only for $\chi_0\gtrsim 10^3$ at the value of $\xi_0=10$ under consideration.

Other two important limiting expressions can be obtained. In the case of a weak plane wave, i.e., for  $|\pmb{\xi}_{\perp}(\varphi)|\ll 1$, the result for the total probability is obtained from the corresponding weak-field limits $I_{1,w}$, $I_{2,w}$, and $I_{3,w}$ of the integrals $I_1$, $I_2$, and $I_3$, respectively. By setting $\Delta\Psi=\Psi-\Psi_0$ and by noticing that the integrals $I_2$ and $I_3$ are already multiplied by a quadratic function in the dimensionless field amplitude, we obtain
\begin{align}
I_{1,w}&=\Delta\Psi\text{Re}[(1-i\Psi_0)e^{-i\Psi_0}\Gamma(0,-i\Psi_0)-1],\\
I_{2,w}&=\text{Im}\,e^{-i\Psi_0}\Gamma(0,-i\Psi_0),\\
I_{3,w}&=\frac{\Psi_0}{2}-\frac{\Psi_0^2}{2}\text{Im}\,e^{-i\Psi_0}\Gamma(0,-i\Psi_0).
\end{align}
Finally, the opposite limit $|\pmb{\xi}_{\perp}(\varphi)|\gg 1$ corresponds to the locally-constant field approximation (LCFA) because the largest contribution of the integral over $\varphi_-$ comes from the region $|\varphi_-|\ll 1$, such that one can approximate
\begin{align}
\pmb{\xi}_{\perp}\left(\varphi_++\frac{\varphi_-}{2}\right)-\pmb{\xi}_{\perp}\left(\varphi_+-\frac{\varphi_-}{2}\right)&\approx \pmb{\xi}'_{\perp}(\varphi_+)\varphi_-,\\
\int_{-\varphi_-/2}^{\varphi_-/2}d\tilde{\varphi}\,\pmb{\xi}^2_{\perp}(\varphi_++\tilde{\varphi})-\frac{1}{\varphi_-}\bigg[\int_{-\varphi_-/2}^{\varphi_-/2}d\tilde{\varphi}\,\pmb{\xi}_{\perp}(\varphi_++\tilde{\varphi})\bigg]^2&\approx \frac{1}{12}\pmb{\xi}^{\prime\,2}_{\perp}(\varphi_+)\varphi_-^3
\end{align}
where the prime denotes the derivative with respect to $\varphi_+$. In this case, it is useful to go back to Eq. (\ref{P}) and take there the integral over $\varphi_-$. One then obtains the probability $P_{\text{LCFA}}$ within the LCFA in the form
\begin{equation}
\begin{split}
P_{\text{LCFA}}&=\frac{\alpha}{\sqrt{3}\pi}\frac{\xi_0}{\chi_0}\int d\varphi_+\int_0^{\infty}\frac{du}{(1+u)^2}\Bigg[\left(1+u+\frac{1}{1+u}\right)\text{K}_{2/3}\left(\frac{2u}{3\chi(\varphi_+)}\right)\\
&\quad\left.-\int_{\frac{2u}{3\chi(\varphi_+)}}^{\infty}dz\,\text{K}_{1/3}(z)\right],
\end{split}
\end{equation}
with $\chi(\varphi)=\chi_0|\pmb{\xi}'_{\perp}(\varphi)|/\xi_0$ and with $\text{K}_{\nu}(\cdot)$ denoting the modified Bessel function of order $\nu$ \cite{NIST_b_2010}, which coincides with the result in, e.g., Ref. \cite{Baier_b_1998}.

\subsection{A note on the difference between the limit $k_-\to 0$ and the limit $\omega\to 0$}
Before proceeding, we would like to address a general aspect of the infrared limit of the probability of nonlinear Compton scattering: The limit $k_-\to 0$ vs the limit $\omega\to 0$. This point has been already discussed in Ref. \cite{Di_Piazza_2018}, where it has been stressed that the two limits are clearly different, as it testified by the fact that the limit of $dP/dk_-$ ($dP/d\omega$) for $k_-\to 0$ ($\omega\to 0$) tends to a non-zero constant (to zero) in the case of non-unipolar fields. Whereas the limit $\lim_{\omega\to 0}dP/d\omega$ is a true soft limit for which the emitted photon four-momentum vanishes, the same is not true for the limit $\lim_{k_-\to 0}dP/dk_-$. Since the differential probability $dP/dk_-$ is obtained after integrating over the transverse photon momentum $\bm{k}_{\perp}$, the limit $\lim_{k_-\to 0}dP/dk_-$ covers not only the soft region (for vanishing values of $\bm{k}_{\perp}$) but also the region $|k_{\parallel}|\to \infty$ (for non-vanishing values of $\bm{k}_{\perp}$), which is unrelated to the soft limit. Since $\omega=(\bm{k}^2_{\perp}/k_-+k_-)/2$, the soft limit $\omega\to 0$ corresponds in light-cone coordinates to taking first the limit $|\bm{k}_{\perp}|\to 0$ and then the limit $k_-\to 0$.

\subsection{Interpretation of the infrared divergence in terms of the formation length}
\label{Subsec. Interpretation of the infrared divergence in terms of the formation length}
As we have already pointed out, in the case of a unipolar field, the total probability $P$ in Eq. (\ref{P}) is logarithmically divergent. The divergence is known to arise because of the emissions of photons with lower-and-lower energy, whose differential number per unit energy linearly diverges (which implies that the total emitted energy is finite). Since the infrared divergence also occurs in the classical limit and for the same reason, we consider the classical limit $P_c$ in Eq. (\ref{P_c}) of the probability $P$, which is easier to analyze. We also assume that in the non-unipolar case the integrals from minus infinity to plus infinity of $|\pmb{\xi}_{\perp}(\varphi)|$ and $\pmb{\xi}^2_{\perp}(\varphi)$ are finite.

Here, we will interpret the infrared divergence in terms of the formation length of photon emission. We recall that the formation length is the region of integration over the variable $\varphi_-$ where the corresponding integral is formed, i.e., has approximately the same value as when it is performed from minus infinity to plus infinity \cite{Ritus_1985,Baier_2005,Di_Piazza_2018}. As we have mentioned, the integral diverges due to the contribution of photons with lower-and-lower energy. It is physically reasonable to expect that photons with lower-and-lower energy, i.e., larger-and-larger wavelength, have larger-and-larger formation length. Thus, since the integrand is symmetric under the exchange $\varphi_-\to-\varphi_-$, we first write it as twice the integral taken for $\varphi_-\ge 0$ and we look at the region of $\varphi_-\gg 1$ with fixed $\varphi_+$. It is convenient to pass to the variables $\varphi=\Phi\sin(\vartheta)$ and $\varphi'=\Phi\cos(\vartheta)$ so that
\begin{equation}
\begin{split}
P_c&=\frac{2\alpha}{\pi}\int_0^{\infty}\frac{d\Phi}{\Phi}\int_{\pi/4}^{5\pi/4}\frac{d\vartheta}{[\sin(\vartheta)-\cos(\vartheta)]^2}\Bigg\{\frac{1}{2}\left[\pmb{\xi}_{\perp}\left(\Phi\sin(\vartheta)\right)-\pmb{\xi}_{\perp}\left(\Phi\cos(\vartheta)\right)\right]^2\\
&\quad\left.-\frac{1}{\sin(\vartheta)-\cos(\vartheta)}\int_{\cos(\vartheta)}^{\sin(\vartheta)}d\rho\,\pmb{\xi}^2_{\perp}(\Phi\rho)+\bigg[\frac{1}{\sin(\vartheta)-\cos(\vartheta)}\int_{\cos(\vartheta)}^{\sin(\vartheta)}d\rho\,\pmb{\xi}_{\perp}(\Phi\rho)\bigg]^2\right\}\\
&\quad\times\left\{1+\frac{1}{\sin(\vartheta)-\cos(\vartheta)}\int_{\cos(\vartheta)}^{\sin(\vartheta)}d\rho\,\pmb{\xi}^2_{\perp}(\Phi\rho)-\bigg[\frac{1}{\sin(\vartheta)-\cos(\vartheta)}\int_{\cos(\vartheta)}^{\sin(\vartheta)}d\rho\,\pmb{\xi}_{\perp}(\Phi\rho)\bigg]^2\right\}^{-1}.
\end{split}
\end{equation}
Now, the regions of $\vartheta$ close to $\pi/4$ or $5\pi/4$ at fixed $\Phi$ correspond to $\varphi_-\ll 1$ and we know from the LCFA that there the integral converges. The ``central'' region $\vartheta\approx 3\pi/4$, corresponding to $|\varphi_+|\ll 1$, needs to be studied in detail. In this region, it is $\sin(\vartheta)\approx 1/\sqrt{2}$ and $\cos(\vartheta)\approx -1/\sqrt{2}$. If $\lim_{\varphi\to\infty}\pmb{\xi}_{\perp}(\varphi)=\bm{0}$, one can easily see that all terms in the numerator of the integrand in $P_c$ vanish in that limit and then the integral converges. However, if $\lim_{\varphi\to\infty}\pmb{\xi}_{\perp}(\varphi)\neq \bm{0}$, in order to take the correct limit, it is necessary to split each integral into two parts where $\rho>0$ and $\rho<0$, respectively. One can then easily obtain that
\begin{equation}
\begin{split}
\lim_{\Phi\to\infty}&\Bigg\{\frac{1}{2}\left[\pmb{\xi}_{\perp}\left(\Phi\sin(\vartheta)\right)-\pmb{\xi}_{\perp}\left(\Phi\cos(\vartheta)\right)\right]^2\\
&\quad\left.-\frac{1}{\sin(\vartheta)-\cos(\vartheta)}\int_{\cos(\vartheta)}^{\sin(\vartheta)}d\rho\,\pmb{\xi}^2_{\perp}(\Phi\rho)+\bigg[\frac{1}{\sin(\vartheta)-\cos(\vartheta)}\int_{\cos(\vartheta)}^{\sin(\vartheta)}d\rho\,\pmb{\xi}_{\perp}(\Phi\rho)\bigg]^2\right\}\\
&\quad\times\left\{1+\frac{1}{\sin(\vartheta)-\cos(\vartheta)}\int_{\cos(\vartheta)}^{\sin(\vartheta)}d\rho\,\pmb{\xi}^2_{\perp}(\Phi\rho)-\bigg[\frac{1}{\sin(\vartheta)-\cos(\vartheta)}\int_{\cos(\vartheta)}^{\sin(\vartheta)}d\rho\,\pmb{\xi}_{\perp}(\Phi\rho)\bigg]^2\right\}^{-1}\\
&=\frac{[\pmb{\xi}_{\perp}(\infty)-\pmb{\xi}_{\perp}(-\infty)]^2}{4+[\pmb{\xi}_{\perp}(\infty)-\pmb{\xi}_{\perp}(-\infty)]^2},
\end{split}
\end{equation}
where for the sake of symmetry we kept the quantity $\pmb{\xi}_{\perp}(-\infty)$ even though it has been assumed to be vanishing here. The fact that this limit is nonzero lets us conclude that the integral in $\Phi$ is logarithmically divergent.

\subsection{A note on the inclusion of radiation-reaction effects}
\label{Subsec. A note on the inclusion of radiation-reaction effects}
The classical spectral intensity of emission, i.e., the energy radiated per unit of angular frequency in an external electromagnetic field corresponds to the classical limit of the average energy emitted via single-photon emission in that external field. This can be explicitly shown in the case of an arbitrary plane wave and also within the quasiclassical approximation based either on Baier's method \cite{Baier_b_1998} or using the quasiclassical wave functions \cite{Di_Piazza_2021}. However, the classical expression of the spectral intensity of radiation is valid for an arbitrary trajectory of the electron and therefore it can also be used when radiation-reaction effects are taken into account in the determination of the trajectory. We have exploited this fact in Refs. \cite{Di_Piazza_2021_b} and  \cite{Di_Piazza_2018_b} to compute the classical emission spectrum of an electron in an arbitrary plane wave including radiation-reaction effects and to study its infrared limit, respectively (see also Ref. \cite{Heinzl_2021} for the computation of the emission spectrum including radiation reaction within the LCFA). Now, at least within the LCFA the quantum emission spectra including radiation reaction tend to the classical expression obtained by using the electron trajectory according to the Landau-Lifshitz equation \cite{Landau_b_2_1975} in the general formula of the emission spectrum \cite{Di_Piazza_2010,Torgrimsson_2021}. This result has been equivalently shown via the kinetic approach \cite{Elkina_2011,Neitz_2013}. This is particularly transparent in the latter approach, where one sees that in the ultrarelativistic limit, the photon yield is still given by the integral of the single-photon emission probability multiplied by the electron distribution function, which in turn is determined by the Liouville equation with the force given by the Landau-Lifshitz force in the same limit \cite{Elkina_2011,Neitz_2013}. 

The above considerations hint to the fact that also when including radiation-reaction effects the classical limit of the differential emission photon yield is given by $dP_c/d^3\bm{k}=\omega^{-1}d\mathcal{E}_c/d^3\bm{k}$, with $d\mathcal{E}_c/d^3\bm{k}$ given in Eq. (\ref{dE_dk}) but with the electron's trajectory being determined via the Landau-Lifshitz equation. In the case of a plane wave, the quantity $dP_c/d^3\bm{k}$ is conveniently expressed in terms of a double integral over the plane-wave phase and then the exact solution of the Landau-Lifshitz equation found in Ref. \cite{Di_Piazza_2008} can be used to determine the differential photon yield. It should be observed that when radiation-reaction effects are included, even if the field is not unipolar, the final momentum of the electron differs from the initial one and, as a consequence, the total radiation yield logarithmically diverges (after it has been regularized by explicitly subtracting the corresponding free-field expression). Thus, in this case it is physically more relevant to compute the number of photons emitted with an energy larger than a given $\omega_m$. Since the procedure is the same, we consider the case of an electron moving on an arbitrary trajectory characterized by time-dependent position $\bm{r}(t)$ and the four-velocity $u^{\mu}(t)=(\gamma(t),\bm{u}(t))=\gamma(t)(1,\bm{v}(t))$ (initial values at, say, $t=0$ given by $\bm{r}_0$ and $u^{\mu}_0=(\gamma_0,\bm{u}_0)=\gamma_0(1,\bm{v}_0))$. Starting from Eq. (\ref{dE_dk}), the probability per unit of photon energy $\omega$ and of solid angle $\Omega$ can be written as (see also Ref. \cite{Baier_b_1998})
\begin{equation}
\label{dP_c_dodO}
\frac{dP_c}{d\omega d\Omega}=-\frac{\alpha\omega}{4\pi^2}\text{Re}\int \frac{dtdt'}{\gamma(t)\gamma(t')}\,\left(1-\frac{(\Delta u)^2}{2}\right)e^{i\omega(\Delta t-\bm{n}\cdot\Delta \bm{r})},
\end{equation}
where $\Delta t=t-t'$, $\Delta \bm{r}=\bm{r}(t)-\bm{r}(t')$, $\Delta u^{\mu}=u^{\mu}(t)-u^{\mu}(t')$ and where the solid angle corresponds to the emission direction $\bm{n}=\bm{k}/\omega$. This expression can explicitly regularized by subtracting the (vanishing) value for a vanishing external field:
\begin{equation}
\frac{dP_c}{d\omega d\Omega}=-\frac{\alpha\omega}{4\pi^2}\text{Re}\int dtdt'\,\left[\frac{e^{i\omega(\Delta t-\bm{n}\cdot\Delta \bm{r})}}{\gamma(t)\gamma(t')}-\frac{e^{i\omega(1-\bm{n}\cdot\bm{v}_0)\Delta t}}{\gamma_0^2}-\frac{(\Delta u)^2}{2}\frac{e^{i\omega(\Delta t-\bm{n}\cdot\Delta \bm{r})}}{\gamma(t)\gamma(t')}\right].
\end{equation}
The integrals over the solid angle can be computed by assuming that the $z$ axis is along the vector multiplying the unit vector $\bm{n}$ in the exponential in each term. In this way, one obtains
\begin{equation}
\begin{split}
\frac{dP_c}{d\omega}=-\frac{\alpha}{2\pi}\text{Im}\int dtdt'\,e^{i\omega\Delta t}\left[\left(1-\frac{(\Delta u)^2}{2}\right)\frac{e^{i\omega|\Delta \bm{r}|}-e^{-i\omega|\Delta \bm{r}|}}{\gamma(t)\gamma(t')|\Delta \bm{r}|}-\frac{e^{i\omega|\bm{v}_0\Delta t|}-e^{-i\omega|\bm{v}_0\Delta t|}}{\gamma_0^2|\bm{v}_0\Delta t|}\right].
\end{split}
\end{equation}
Finally, the integral over $\omega$ is taken assuming that the phases feature infinitesimal imaginary parts such that the integrals converge for $\omega\to\infty$ and the final result is
\begin{equation}
\begin{split}
P_c(\omega_m)&=\frac{\alpha}{\pi}\int dtdt'\,\left\{\left(1-\frac{(\Delta u)^2}{2}\right)\right.\\
&\quad\times\frac{\Delta t\sin(\omega_m\Delta t)\sin(\omega_m|\Delta \bm{r}|)+|\Delta \bm{r}|\cos(\omega_m\Delta t)\cos(\omega_m|\Delta \bm{r}|)}{\gamma(t)\gamma(t')|\Delta \bm{r}|[(\Delta t)^2-(\Delta \bm{r})^2]}\\
&\quad\left.-\frac{\Delta t\sin(\omega_m\Delta t)\sin(\omega_m|\bm{v}_0\Delta t|)+|\bm{v}_0\Delta t|\cos(\omega_m\Delta t)\cos(\omega_m|\bm{v}_0\Delta t|)}{|\bm{v}_0\Delta t|(\Delta t)^2}\right\}.
\end{split}
\end{equation}
It can be easily seen that the integrand is regular for $\Delta t\to 0$ and that $|\Delta t|\ge |\Delta \bm{r}|$, with the equality occurring only for $\Delta t=0$. The inequality can be proved by noticing that
\begin{equation}
|\Delta \bm{r}|=|\bm{r}(t)-\bm{r}(t')|=\left|\int_{t'}^td\tilde{t}\,\bm{v}(\tilde{t})\right|\le \left|\int_{t'}^td\tilde{t}\,|\bm{v}(\tilde{t})|\right|\le v_M|\Delta t|\le |\Delta t|,
\end{equation}
with $v_M<1$ being the maximal value of the modulus of the electron velocity between $t'$ and $t$ and the equality in the last step occurring only for $\Delta t=0$.

In the specific case of an electron moving in a plane wave the yield $P_c(\omega_m)$ can be expressed as integrals over the plane-wave light-cone coordinate $\phi$ by means of the following transformation formulas:
\begin{align}
\int dtdt'&\rightarrow \int d\phi d\phi'\frac{\varepsilon(\phi)\varepsilon(\phi')}{\pi_-(\phi)\pi_-(\phi')},\\
\Delta t=\int_{t'}^td\tilde{t}&\rightarrow \int_{\phi'}^{\phi}d\tilde{\phi}\frac{\varepsilon(\tilde{\phi})}{\pi_-(\tilde{\phi})},\\
\Delta \bm{r}=\int_{t'}^td\tilde{t}\,\bm{v}(\tilde{t})&\rightarrow \int_{\phi'}^{\phi}d\tilde{\phi}\frac{\bm{\pi}(\tilde{\phi})}{\pi_-(\tilde{\phi})},
\end{align}
where $\pi^{\mu}(\phi)=(\varepsilon(\phi),\bm{\pi}(\phi))$ is the electron four-momentum, and then use the solution in Ref. \cite{Di_Piazza_2008_a} for the electron four-momentum including radiation-reaction effects (recall that the light-cone momentum $\pi_-(\phi)$ is not a constant of motion in this case).

\section{Classical angular distribution of radiation}
\label{Sec. Classical angular distribution of radiation}
In laser field-electron beam experiments it is usually not possible to detect photons at arbitrary emission angles. This is especially the case of low-energy photons as their angular distribution is typically broader than that of high-energy photons. Therefore, the computation of the yield of photons emitted with angular frequencies larger than a fixed $\omega_m$ and at a fixed solid angle is of interest. We start from Eq. (\ref{dP_c_dodO}) and we rewrite it in the convenient form
\begin{equation}
\frac{dP_c}{d\omega d\Omega}=\frac{\alpha}{4\pi^2\omega}\int dtdt'\bm{F}(t)\cdot\bm{F}(t')e^{i\omega\int_{t'}^td\tilde{t}[1-\bm{n}\cdot\bm{v}(\tilde{t})]},
\end{equation}
with
\begin{equation}
\bm{F}(t)=\frac{\bm{n}\times\{[\bm{n}-\bm{v}(t)]\times\dot{\bm{v}}(t)\}}{[1-\bm{n}\cdot\bm{v}(t)]^2},
\end{equation}
which is actually the original expression that one obtains in classical electrodynamics starting from the Li\'{e}nard-Wiechert potentials \cite{Jackson_b_1975} (the dot indicates the derivative with respect to time) and which is convenient for numerical evaluations.

The integral over the angular frequency from a lower bound $\omega_m$ to infinity can expressed in terms of an incomplete Gamma function as
\begin{equation}
\begin{split}
\frac{dP_c}{d\Omega}&=\frac{\alpha}{2\pi^2}\text{Re}\int dt_+\int_0^{\infty}dt_-\bm{F}\left(t_++\frac{t_-}{2}\right)\cdot\bm{F}\left(t_+-\frac{t_-}{2}\right)\\
&\quad\times\Gamma\left(0,-i\omega_m\int_{t_+-t_-/2}^{t_++t_-/2}d\tilde{t}[1-\bm{n}\cdot\bm{v}(\tilde{t})]\right),
\end{split}
\end{equation}
where the variables $t_{\pm}$ have been defined analogously as the variables $\varphi_{\pm}$ and where taking the integral over $t_-$ from zero to infinity ensures that the time-integral in the incomplete Gamma function is non-negative. Since we are interested in collecting down to photons with relatively low frequency, we consider the case where (recall that $t_-\ge 0$)
\begin{equation}
\label{Cond}
\omega_m\int_{t_+-t_-/2}^{t_++t_-/2}d\tilde{t}[1-\bm{n}\cdot\bm{v}(\tilde{t})]\ll 1
\end{equation}
and, using the known expansion of the incomplete Gamma function \cite{NIST_b_2010}, we obtain
\begin{equation}
\label{dP_dO}
\begin{split}
\frac{dP_c}{d\Omega}&=-\frac{\alpha}{2\pi^2}\int dt_+\int_0^{\infty}dt_-\bm{F}\left(t_++\frac{t_-}{2}\right)\cdot\bm{F}\left(t_+-\frac{t_-}{2}\right)\\
&\quad\times\left\{\log\left(\omega_m\int_{t_+-t_-/2}^{t_++t_-/2}d\tilde{t}[1-\bm{n}\cdot\bm{v}(\tilde{t})]\right)+\gamma_E\right\},\\
&=\frac{\alpha}{4\pi^2}\int dtdt'\bm{F}(t)\cdot\bm{F}(t')\left\{\log\left(\frac{1}{\omega_m\left|\int_{t'}^td\tilde{t}[1-\bm{n}\cdot\bm{v}(\tilde{t})]\right|}\right)-\gamma_E\right\},
\end{split}
\end{equation}
where $\gamma_E=0.577\ldots$ is the Euler constant. Note that, as it should be, the leading term proportional to the logarithm of $\omega_m$ (divided by a constant with the same dimensions) would vanish if the initial and final four-momenta of the electron coincide because the function $\bm{F}(t)$ is a total time derivative \cite{Jackson_b_1975}.

Before concluding this section, we provide a simpler parametric condition of validity of the condition in Eq. (\ref{Cond}) in the case of a plane wave. By ignoring for simplicity radiation-reaction effects, i.e., the four-momentum of the photon emitted with the minimal energy $\omega_m$, we have that (see also Eq. (\ref{dP_d^3k}) and recall that $\chi_0/\xi_0=(k_Lp)/m^2$)
\begin{equation}
\omega_m\int_{t_+-t_-/2}^{t_++t_-/2}d\tilde{t}[1-\bm{n}\cdot\bm{v}(\tilde{t})]=\frac{k_{m,-}}{2p_-}\frac{\xi_0}{\chi_0}\int_{\varphi}^{\varphi'}d\tilde{\varphi}\,\left\{1+\left[\frac{\bm{p}_{\perp}}{m}-\frac{p_-}{k_{m,-}}\frac{\bm{k}_{m,\perp}}{m}-\bm{\xi}_{\perp}(\tilde{\varphi})\right]^2\right\}.
\end{equation}
This expression leads to the parametric condition $k_{m,-}/p_-\ll\chi_0/\xi_0^3$, which is the same condition of violation of the LCFA in the infrared part of the spectrum \cite{Di_Piazza_2018}. As we will also see below, a more accurate numerical estimate is obtained by requiring that the phase in the whole integration region is much smaller than unity. If the laser field lasts for a total phase $\Phi_L$, then the condition reads $k_{m,-}/p_-\ll 2\chi_0/(\Phi_L\xi_0^3)$.

\section{Nonlinear Compton scattering probability in a unipolar plane-wave field}
\label{Sec. Nonlinear Compton scattering probability in an unipolar plane-wave field}
As we have seen in the previous section, the classical expression $dP_c/d^3\bm{k}$ of the differential radiation yield does not explicitly depend on whether the plane wave is unipolar or not. The fact that the total yield diverges or not is implicitly encoded in the electron trajectory. One can ask the question whether this is also the case in strong-field QED, i.e., when quantum effects are important. 

The probability of nonlinear Compton scattering in a unipolar plane-wave field has been studied in Ref. \cite{Dinu_2012}. Since, $\bm{A}_{\perp}(\infty)\neq \bm{0}$, the Volkov out-states have been first computed and then used to determine the emission probability. We first re-derive the Volkov out-state in a different way as in Ref. \cite{Dinu_2012}, then we will compute the corresponding probability of nonlinear Compton scattering, and finally we will show that all the occurrences of $\bm{A}_{\perp}(\infty)$ cancel out in the probability. In this way, analogously as in classical electrodynamics, the logarithmic divergence is implicitly encoded in the behavior of the vector potential $\bm{A}_{\perp}(\varphi)$ at infinity.

In order to find the Volkov out-states, we write the Dirac equation in the plane wave as
\begin{equation}
\{\gamma^{\mu}[i\partial_{\mu}-eA_{\mu}(\phi)]-m\}\psi=\{\gamma^{\mu}[i\partial_{\mu}-e\delta A_{\mu}(\phi)-eA_{\mu}(\infty)]-m\}\psi=0
\end{equation}
where $\gamma^{\mu}$ are the Dirac matrices and $\delta A^{\mu}(\phi)=A^{\mu}(\phi)-A^{\mu}(\infty)$. Since $\delta A^{\mu}(\phi)$ vanishes at $\phi\to \infty$ and since the extra term proportional to $A^{\mu}(\infty)$ is a constant, which can be eliminated by a gauge-like transformation, the Volkov out-state $\psi^{(\text{out})}_s(x;\bm{p})$ with asymptotic on-shell four-momentum $p^{\mu}=(\varepsilon,\bm{p})$ and spin quantum number $s$ reads
\begin{equation}
\psi^{(\text{out})}_s(x;\bm{p})=e^{i\left\{-(px)-e(A(\infty)x)-\int_{\infty}^{\phi}d\phi'\left[\frac{e(p\delta A)}{p_-}-\frac{e^2(\delta A)^2}{2p_-}\right]\right\}}\left[1+e\frac{\hat{n}_L\hat{\delta A}(\phi)}{2p_-}\right]u_s(\bm{p}),
\end{equation}
where the hat on a four-vector indicates the contraction with the gamma matrices and where $u_s(\bm{p})$ is the corresponding constant spinor, in agreement with Ref. \cite{Dinu_2012}.

Now, the probability amplitude $S_{fi}$ that an electron with initial quantum numbers $\bm{p}$ (energy $\varepsilon$) and $s$ emits a photon with quantum numbers $\bm{k}$ (energy $\omega$) and polarization $l$ going into the state characterized by the quantum numbers $\bm{p}'$ (energy $\varepsilon'$) and $s'$ is (apart from an inconsequential constant phase)
\begin{equation}
\begin{split} \label{Eq. S_fi}
S_{fi}&=-ie(2\pi)^3\delta(p'_-+k_--p_-)\delta^2(\bm{p}'_{\perp}+e\bm{A}_{\perp}(\infty)+\bm{k}_{\perp}-\bm{p}_{\perp})\\
&\quad\times\int d\phi\bar{u}_{s'}(\bm{p}')\left[1-e\frac{\hat{n}_L\hat{\delta A}(\phi)}{2p'_-}\right]\hat{e}^*_l(\bm{k})\left[1+e\frac{\hat{n}_L\hat{A}(\phi)}{2p_-}\right]\\
&\quad\times e^{i\left\{(p'_++k_+-p_+)\phi-\int_0^{\phi}d\phi'\left[-e\frac{\bm{p}_{\perp}\cdot \bm{A}_{\perp}}{p_-}+e\frac{\bm{p}'_{\perp}\cdot \delta\bm{A}_{\perp}}{p'_-}+\frac{e^2\bm{A}^2_{\perp}}{2p_-}-\frac{e^2\delta\bm{A}^2_{\perp}}{2p'_-}\right]\right\}},
\end{split}
\end{equation}
where we have already exploited the symmetries of the background plane wave, where the bar on a spinor indicates its Dirac conjugated, and where $e_l^{\mu}(\bm{k})$ is the polarization four-vector of the emitted photon. By using the same procedure as, for example, in Ref. \cite{Di_Piazza_2018}, one obtains the differential probability averaged (summed) over the initial (final) discrete quantum numbers  in the form
\begin{equation}
\begin{split}
\frac{dP}{d^3\bm{k}}&=-\frac{\alpha}{32\pi^2\omega p_-p'_-}\int d\phi d\phi'\,\text{Tr}\left\{(\hat{p}+m)\left[1-e\frac{\hat{n}_L\hat{ A}(\phi')}{2p_-}\right]\gamma^{\mu}\left[1+e\frac{\hat{n}_L\hat{\delta A}(\phi')}{2p'_-}\right]\right.\\
&\quad\times\left.(\hat{p}'+m)\left[1-e\frac{\hat{n}_L\hat{\delta A}(\phi)}{2p'_-}\right]\gamma_{\mu}\left[1+e\frac{\hat{n}_L\hat{A}(\phi)}{2p_-}\right]\right\}\\
&\quad\times e^{i\left\{\left(\frac{m^2+\bm{p}_{\perp}^{\prime\,2}}{2p'_-}+\frac{\bm{k}^2_{\perp}}{2k_-}-\frac{m^2+\bm{p}^2_{\perp}}{2p_-}\right)(\phi-\phi')-\int_{\phi'}^{\phi}d\tilde{\phi}\left[-e\frac{\bm{p}_{\perp}\cdot \bm{A}_{\perp}}{p_-}+e\frac{\bm{p}'_{\perp}\cdot \delta\bm{A}_{\perp}}{p'_-}+\frac{e^2\bm{A}^2_{\perp}}{2p_-}-\frac{e^2\delta\bm{A}^2_{\perp}}{2p'_-}\right]\right\}}.
\end{split}
\end{equation}
Starting from the phase, one can show that all the occurrences of $\bm{A}_{\perp}(\infty)$ cancel out by using the conservation laws $p'_-=p_--k_-$ and $\bm{p}'_{\perp}=\bm{p}_{\perp}-\bm{k}_{\perp}-e\bm{A}_{\perp}(\infty)$ and by using the definition of $\delta\bm{A}_{\perp}(\phi)$ such that
\begin{equation}
\begin{split}
&e^{i\left\{\left(\frac{m^2+\bm{p}_{\perp}^{\prime\,2}}{2p'_-}+\frac{\bm{k}^2_{\perp}}{2k_-}-\frac{m^2+\bm{p}^2_{\perp}}{2p_-}\right)(\phi-\phi')-\int_{\phi'}^{\phi}d\tilde{\phi}\left[-e\frac{\bm{p}_{\perp}\cdot \bm{A}_{\perp}}{p_-}+e\frac{\bm{p}'_{\perp}\cdot \delta\bm{A}_{\perp}}{p'_-}+\frac{e^2\bm{A}^2_{\perp}}{2p_-}-\frac{e^2\delta\bm{A}^2_{\perp}}{2p'_-}\right]\right\}}\\
&\quad=e^{i\left\{\left[\frac{m^2+(\bm{p}_{\perp}-\bm{k}_{\perp})^2}{2p'_-}+\frac{\bm{k}^2_{\perp}}{2k_-}-\frac{m^2+\bm{p}^2_{\perp}}{2p_-}\right](\phi-\phi')-\int_{\phi'}^{\phi}d\tilde{\phi}\left[-e\frac{\bm{p}_{\perp}\cdot \bm{A}_{\perp}}{p_-}+e\frac{(\bm{p}_{\perp}-\bm{k}_{\perp})\cdot \bm{A}_{\perp}}{p'_-}+\frac{e^2\bm{A}^2_{\perp}}{2p_-}-\frac{e^2\bm{A}^2_{\perp}}{2p'_-}\right]\right\}}.
\end{split}
\end{equation}
This can also be proven by observing that the phase in Eq. \eqref{Eq. S_fi} can be written in the form 
\begin{equation}
\begin{split}
&e^{i\left\{(p'_++k_+-p_+)\phi-\int_0^{\phi}d\phi'\left[-e\frac{\bm{p}_{\perp}\cdot \bm{A}_{\perp}}{p_-}+e\frac{\bm{p}'_{\perp}\cdot \delta\bm{A}_{\perp}}{p'_-}+\frac{e^2\bm{A}^2_{\perp}}{2p_-}-\frac{e^2\delta\bm{A}^2_{\perp}}{2p'_-}\right]\right\}} 
\\& \quad = 
e^{i
\int_0^{\phi}d\phi'\left[
k_+ +m^2\left(\frac{1}{2p'_-}-\frac{1}{2p_-}\right)
- \frac{1}{2p_-}
\left(\bm{p}_{\perp} - e \bm{A}_\perp     \right) ^2
+ \frac{1}{2p'_-}
\left(\bm{p'}_{\perp} - e \delta \bm{A}_\perp     \right) ^2
\right]
},
\end{split}
\end{equation}
which clarifies that, due to the transverse momentum conservation, adding a dc component to the plane-wave field simply results in shifting the electron final transverse momentum: $\bm{p'}_{\perp} - e \delta \bm{A}_\perp(\phi)  = \bm{p}_{\perp} - \bm{k}_\perp - e  \bm{A}_\perp(\phi) $.

Passing now to the trace, we see that the only matrix containing $\bm{A}_{\perp}(\infty)$ is
\begin{equation}
\begin{split}
&\left[1+e\frac{\hat{n}_L\hat{\delta A}(\phi')}{2p'_-}\right](\hat{p}'+m)\left[1-e\frac{\hat{n}_L\hat{\delta A}(\phi)}{2p'_-}\right]\\
&\quad=\left[1+e\frac{\hat{n}_L\hat{A}(\phi')}{2p'_-}\right]\left[1-e\frac{\hat{n}_L\hat{A}(\infty)}{2p'_-}\right](\hat{p}'+m)\left[1+e\frac{\hat{n}_L\hat{A}(\infty)}{2p'_-}\right]\left[1-e\frac{\hat{n}_L\hat{A}(\phi)}{2p'_-}\right]\\
&\quad=\left[1+e\frac{\hat{n}_L\hat{A}(\phi')}{2p'_-}\right](\hat{P}'+m)\left[1-e\frac{\hat{n}_L\hat{A}(\phi)}{2p'_-}\right]
\end{split}
\end{equation}
where
\begin{equation}
P^{\prime\,\mu}=p^{\prime\,\mu}+eA^{\mu}(\infty)-\frac{e(p'A(\infty))}{p'_-}n_L^{\mu}-\frac{e^2A^2(\infty)}{2p'_-}n_L^{\mu}.
\end{equation}
The fact that $P^{\prime\,\mu}$ does not depend on $A^{\mu}(\infty)$ can be conveniently shown via its light-cone components, which can be simplified again using the conservation laws:
\begin{align}
P'_-&=p'_-,\\
\bm{P}'_{\perp}&=\bm{p}'_{\perp}+e\bm{A}_{\perp}(\infty)=\bm{p}_{\perp}-\bm{k}_{\perp},\\
P'_+&=p'_++\frac{e\bm{p}'_{\perp}\cdot\bm{A}_{\perp}(\infty)}{p'_-}+\frac{e^2\bm{A}^2_{\perp}(\infty)}{2p'_-}=\frac{m^2+(\bm{p}_{\perp}-\bm{k}_{\perp})^2}{2p'_-}.
\end{align}
These results allow one to conclude that, analogously as in the classical case, the differential probability $dP/d^3\bm{k}$ has the same expression independently on whether the field is unipolar or not. The resulting logarithmic divergence in the former case is encoded in the non-vanishing limit of $\bm{A}_{\perp}(\phi)$ for $\phi\to\infty$.

\section{Emitted photon yield in an arbitrary field within the quasiclassical approximation}
\label{Sec. Emitted photon yield in an arbitrary field within the quasiclassical approximation}
In this section, we study the problem of the photon emission yield in a field of virtually arbitrary spacetime structure within the quasiclassical approximation, having in mind the case of a tightly focused laser beam. Our considerations will be based on the method presented in Ref. \cite{Di_Piazza_2021}. The main novelty as compared to Ref. \cite{Di_Piazza_2021} is that we will consider here an incoming electron described by a wave packet (well centered around a given asymptotic momentum) rather than by a state with definite asymptotic momentum. Now, the notation in Ref. \cite{Di_Piazza_2021} clashes for some symbols with what we have used so far here. Thus, in order to avoid confusion, we will present the derivation of the emission probability per unit of photon energy and solid angle in the appendix (using the notation of Ref. \cite{Di_Piazza_2021}) and we report here the final result by using the notation of the rest of the paper. If the probability density associated with the initial wave packet is described by the function $|f(t,\bm{x})|^2$ outside the field, we find that the differential probability of emitting a photon (averaged/summed over the initial/final discrete quantum numbers of the electron and the photon) is given by (see, in particular, Eq. (\ref{dP_d^3k_Baier}) and the discussion below it)
\begin{equation}
\label{dP_d^3k_Baier_2}
\begin{split}
\frac{dP}{d\omega d\Omega}&=\int d^3\bm{x}_0|f(t_0,\bm{x}_0)|^2\frac{\alpha}{4\pi^2\omega}\left\{\frac{\varepsilon^{\prime\,2}+\varepsilon^2}{2\varepsilon^2}\left|\int dt\,\bm{F}(t)e^{i\frac{\omega\varepsilon}{\varepsilon'}\int_0^td\tilde{t}[1-\bm{n}\cdot\bm{v}(\tilde{t})]}\right|^2\right.\\
&\quad\left.+\frac{\omega^2 m^2}{2\varepsilon^4}\left|\int dt\,G(t)e^{i\frac{\omega\varepsilon}{\varepsilon'}\int_0^td\tilde{t}[1-\bm{n}\cdot\bm{v}(\tilde{t})]}\right|^2\right\},
\end{split}
\end{equation}
where $t_0$ represents an initial time where the wave packet is outside the external field, where $\varepsilon'=\varepsilon-\omega$, and where
\begin{equation}
G(t)=\frac{\bm{n}\cdot\dot{\bm{v}}(t)}{[1-\bm{n}\cdot\bm{v}(t)]^2}.
\end{equation}
Apart from the average over the initial position of each trajectory determined via the Lorentz equation, one recognizes the spectrum according to Baier's method \cite{Baier_b_1998}, with the additional integration by parts in the time integrals \cite{Wistisen_2014}, which is convenient for numerical evaluations. By indicating as $\langle\rangle$ the average over the initial electron wave function for notational simplicity, the integral over the photon energy can be taken by performing the change of variable $u=\omega/(\varepsilon-\omega)$. The result is again written in terms of the incomplete Gamma function \cite{NIST_b_2010}:
\begin{equation}
\label{dP_dOmega}
\begin{split}
\frac{dP}{d\Omega}&=\frac{\alpha}{4\pi^2}\left\langle\int dt_+\int_0^{\infty}dt_-\left\{[I_{q,1}(u_m)+I_{q,2}(u_m)]\bm{F}\left(t_++\frac{t_-}{2}\right)\cdot\bm{F}\left(t_+-\frac{t_-}{2}\right)\right.\right.\\
&\quad\left.+\frac{m^2}{\varepsilon^2}I_{q,3}(u_m)G\left(t_++\frac{t_-}{2}\right)G\left(t_+-\frac{t_-}{2}\right)\right\}\bigg\rangle,
\end{split}
\end{equation}
where \cite{NIST_b_2010}
\begin{align}
I_{q,1}(u_m)&=\text{Re}\int_{u_m}^{\infty}\frac{du}{u}\frac{e^{iu\varPhi}}{1+u}=\text{Re}[\Gamma(0,-i\varPhi u_m)-e^{-i\varPhi }\Gamma(0,-i\varPhi(1+u_m))],\\
\begin{split}
I_{q,2}(u_m)&=\text{Re}\int_{u_m}^{\infty}\frac{du}{u}\frac{e^{iu\varPhi}}{(1+u)^3}=\text{Re}\left\{\Gamma(0,-i\varPhi u_m)-\left[1+i\frac{\varPhi}{2}+\frac{1}{2(1+u_m)}\right]\frac{e^{iu_m\varPhi}}{1+u_m}\right.\\
&\quad\left.-\left(1+i\varPhi-\frac{\varPhi^2}{2}\right)e^{-i\varPhi}\Gamma(0,-i\varPhi(1+u_m))\right\},
\end{split}\\
\begin{split}
I_{q,3}(u_m)&=\text{Re}\int_{u_m}^{\infty}du\frac{ue^{iu\varPhi}}{(1+u)^3}=\text{Re}\left\{\left[1-i\frac{\varPhi}{2}-\frac{1}{2(1+u_m)}\right]\frac{e^{iu_m\varPhi}}{1+u_m}\right.\\
&\quad\left.+i\varPhi\left(1-i\frac{\varPhi}{2}\right)e^{-i\varPhi}\Gamma(0,-i\varPhi(1+u_m))\right\},
\end{split}
\end{align}
with $u_m=\omega_m/(\varepsilon-\omega_m)$ and
\begin{equation}
\varPhi=\varepsilon\int_{t'}^td\tilde{t}[1-\bm{n}\cdot\bm{v}(\tilde{t})].
\end{equation}

The above formulas can be used to compute the first quantum correction to the classical spectrum. The classical limit is performed as the double limit $u_m\to 0$ and $\varPhi\to\infty$ such that
\begin{equation}
u_m\varPhi\to\omega_m\int_{t'}^td\tilde{t}[1-\bm{n}\cdot\bm{v}(\tilde{t})]
\end{equation}
remains finite. The calculation is tedious but straightforward because it consists in an expansion of the incomplete Gamma functions contained in the integrals $I_{q,1}(u_m)$, $I_{q,2}(u_m)$, and $I_{q,3}(u_m)$. In the limit $u_m|\varPhi|\ll 1$, which we also studied in the classical case, the final result is
\begin{equation}
\label{dP_dOmega_as}
\begin{split}
\frac{dP}{d\Omega}&=\frac{\alpha}{4\pi^2}\int dtdt'\bm{F}(t)\cdot\bm{F}(t')\left\{\log\left(\frac{1}{\omega_m\left|\int_{t'}^td\tilde{t}[1-\bm{n}\cdot\bm{v}(\tilde{t})]\right|}\right)-\gamma_E+\frac{\omega_m}{\varepsilon}\right\}
\end{split}
\end{equation}
to be compared with Eq. (\ref{dP_dO}). This equation shows that the quantum correction, i.e., the last term in the braces is proportional to $\omega_m/\varepsilon$ and as such it is very small in the ultrarelativistic case, where the quasiclassical approximation applies.

\begin{figure}
\begin{center}
\includegraphics[width=0.9\textwidth]{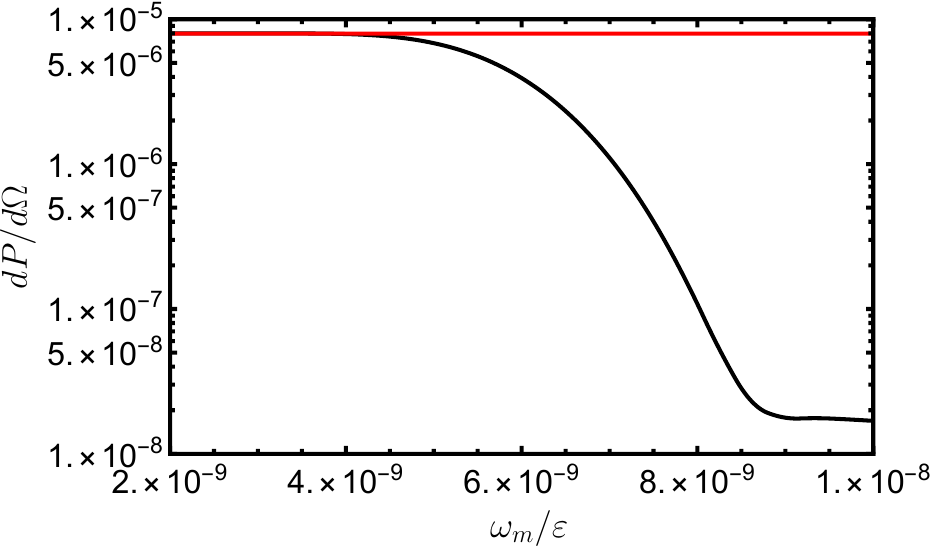}
\end{center}
\caption{Differential probability $dP/d\Omega$ according to Eq. (\ref{dP_dOmega}) (black curve) and the corresponding low-energy asymptotic in Eq. (\ref{dP_dOmega_as}) (red curve) as functions of $\omega_m/\varepsilon$ and for other numerical parameters given in the text.}
\label{dP_dOmega_Plot}
\end{figure}

We conclude this section with a numerical evaluation of Eq. (\ref{dP_dOmega}) and a check on the asymptotic expression in Eq. (\ref{dP_dOmega_as}), even though the quantum correction is too small to be appreciable. For the sake of simplicity, we assume that we can neglect the initial spatial spread of the electron wave packet and that it head-on collides with a Gaussian beam. The electron has an initial energy of $250\;\text{MeV}$ ($\gamma_0\approx 490$) whereas the laser field is linearly polarized, has a central wavelength of $0.8\;\text{$\mu$m}$, the focal waist of $1\;\text{$\mu$m}$, a sin-square temporal profile covering four periods (corresponding to $\Phi_L=8\pi$ and a duration of about $10\;\text{fs}$), and a peak intensity of $1.2\times 10^{22}\;\text{W/cm$^2$}$ \cite{Karsch_2008}, corresponding to $\xi_0\approx 53$ and $\chi_0\approx 0.16$. In Fig. \ref{dP_dOmega_Plot} we show the differential probability of emission as a function of $\omega_m/\varepsilon$ (black curve) and the corresponding asymptotic expression in Eq. (\ref{dP_dOmega_as}). The solid angle corresponds to the direction along the polarization plane of the laser field and at an angle of about $5.8$ degrees from the initial direction of propagation of the electron.
The figure shows that the asymptotic expression works very well for $\omega_m/\varepsilon\lesssim 5\times 10^{-9}$, which is in agreement with the analytical estimation of the validity region according to $\omega_m/\varepsilon\ll 2\chi_0/(\xi_0^3\Phi_L)\approx 8\times 10^{-8}$, given in the previous section. As an additional analytical check, the simulation shows that the final momentum of the electron is $\bm{p}'\approx (0.153137\ldots,0,249.999...)\;\text{MeV}$. By approximating the argument of the logarithm, according to the discussion above, with the constant $4\pi(\omega_m/\varepsilon)(\xi_0^3/\chi_0)$, which is about $2.38\times 10^{-2}$ for $\omega_m/\varepsilon=2\times 10^{-9}$, we can estimate $dP/d\Omega\sim 8.4\times 10^{-6}$, which is in very good agreement with the numerical value $8.0\times 10^{-6}$. The results have been obtained for realistic values of the electron energy and of the laser parameters. Thus, it would be realistic at least in principle to design an experiment to measure the logarithmic behavior of the photon yield in the low-energy regime. By measuring the final electron momentum by means of a spectrometer, one could even determine the constant multiplying the logarithm of the photon energy and have an additional indirect measurement of the photon yield to be used as a consistency check.

\section{Conclusions}
\label{Conclusions}
In this paper we have investigated the yield of photons emitted via nonlinear Thomson and Compton scattering. We have started from the case of a plane-wave background electromagnetic field and we have found an exact analytical formula of the total photon yield in non-unipolar fields, expressed in terms of a double integral over the plane-wave phase. The case of unipolar fields can be treated analogously but the total photon yield diverges logarithmically. In this case, we have determined analytically the photon yield with angular frequencies larger than a minimum value $\omega_m$. Since the two cases of unipolar and non-unipolar fields result in the same expression of the photon yield and the divergence is encoded in the time dependence of the momentum of the electron, we have first shown how this divergence can be connected to the larger-and-larger formation lengths of photons emitted with lower-and-lower energy. 

Furthermore, we have shown that the quantum emission probability in a plane wave has exactly the same expression for unipolar and non-unipolar fields, despite the fact that the former case needs a special treatment in the determination of out-states due to the non-vanishing vector potential at infinity. The latter quantity, in fact, ultimately cancels out in the expression of the probability. 

In the last part of the paper, we passed to the treatment of the more general case of nonlinear Compton scattering in the presence of an arbitrary background field but within the quasiclassical approximation. After deriving (in the appendix) the differential emission probability in the case of an incoming electron wave packet, we have computed the first quantum correction to the angular distribution of photons emitted with energies larger than a minimum value $\omega_m$. We have found that within the quasiclassical approximation, the first quantum correction scales as $\omega_m/\varepsilon$ and it is therefore very small in the ultrarelativistic limit $\varepsilon\gg m$, where the quasiclassical approximation works. By means of a numerical example employing realistic laser and electron parameters we have discussed how the low-energy part of the photon yield could be measured in principle for the first time in a laser-based strong-field QED experiment.

\begin{acknowledgments}
A.D.P. is partially supported by the U.S. National Science Foundation Mid-scale Research Infrastructure Program under Award No. PHY-2329970.

This material is based upon work supported by the U.S. Department of Energy [National Nuclear Security Administration] University of Rochester ``National Inertial Confinement Fusion Program'' under Award Number DE-NA0004144.

This report was prepared as an account of work sponsored by an agency of the United States Government. Neither the United States Government nor any agency thereof, nor any of their employees, makes any warranty, express or implied, or assumes any legal liability or responsibility for the accuracy, completeness, or usefulness of any information, apparatus, product, or process disclosed, or represents that its use would not infringe privately owned rights. Reference herein to any specific commercial product, process, or service by trade name, trademark, manufacturer, or otherwise does not necessarily constitute or imply its endorsement, recommendation, or favoring by the United States Government or any agency thereof. The views and opinions of authors expressed herein do not necessarily state or reflect those of the United States Government or any agency thereof.

\end{acknowledgments}

\appendix
\section{Derivation of the probability within the quasiclassical approximation}
As we have mentioned in the main text, we will derive here the differential probability that an electron emits a photon in an arbitrary background field within the quasiclassical approximation. The physical situation we have in mind is that of a tightly focused laser pulse. Since we will use the same technique as in Ref. \cite{Di_Piazza_2021}, it is convenient to employ here the notation used there. Notice that this notation is employed only in this appendix and the final result is then ``translated'' into the notation used in the rest of the paper.

We assume that the laser field main propagation direction corresponds to the negative $z$ axis and introduce the light-cone coordinates $T=(t+z)/2$, $\bm{x}_{\perp}=(x,y)$, and $\phi=t-z$ for the spacetime point with coordinates $x^{\mu}=(t,\bm{x})=(t,x,y,z)$. The light-cone coordinates of a generic four-vector $v^{\mu}=(v_0,\bm{v})$ are defined as $v_+=(v_0+v_z)/2$, $\bm{v}_{\perp}=(v_x,v_y)$, and $v_-=v_0-v_z$. The same definition applies to the Dirac gamma matrices $\gamma^{\mu}=(\gamma^0,\bm{\gamma})$: $\gamma_+=(\gamma^0+\gamma^3)/2$, $\bm{\gamma}_{\perp}=(\gamma^1,\gamma^2)$, and $\gamma_-=\gamma^0-\gamma^3$. We introduce the four-dimensional quantities: $n^{\mu}=(1,\bm{n})$ and $\tilde{n}^{\mu}=(1,-\bm{n})/2$, with $\bm{n}=(0,0,1)$ and $a_j^{\mu}=(0,\bm{a}_j)$, where $j=1,2$, with $\bm{a}_1=(1,0,0)$ and $\bm{a}_2=(0,1,0)$. We will consider an ultrarelativistic electron (almost) propagating along the positive $z$ direction, and therefore we use/interpret $T$ as the light-cone time and  $\bm{x}_{\text{lc}}=(\bm{x}_{\perp},\phi)$ as the light-cone spatial coordinates. Correspondingly, for a generic on-shell electron four-momentum $p^{\mu}=(\varepsilon,\bm{p})$ ($\varepsilon=\sqrt{m^2+\bm{p}^2}$), we will consider $\bm{p}_{\text{lc}}=(\bm{p}_{\perp},p_+)$ as the three spatial light-cone components of the four-momentum and $p_-=(m^2+\bm{p}_{\perp}^2)/2p_+$ as the time light-cone component. 

Also, we assume that the background electromagnetic field is described by the four-vector potential $A^{\mu}(x)=(V(x),\bm{A}(x))$ satisfying the Lorenz-gauge condition $\partial_{\mu}A^{\mu}(x)=0$ with the additional constraint $A_-(x)=0$, i.e., $A_z(x)=V(x)$ and the asymptotic conditions $\lim_{T\to\pm\infty}A^{\mu}(T,\bm{x}_{\text{lc}})=0$. A fixed value $T_0$ of the light-cone time $T$ is chosen to assign asymptotic conditions of the electron classical trajectory. We assume that the absolute value of $T_0$ is sufficiently large that the background field is negligible, i.e., $A^{\mu}(T_0,\bm{x}_{\text{lc}})=0$.

Unlike in Ref. \cite{Di_Piazza_2021} where we considered the emission of a photon by an electron with a definite asymptotic momentum, we want to study here the more general situation where the incoming electron is described by the wave packet
\begin{equation}
\Psi^{(\text{in})}_s(x)=\int\frac{d^3\bm{p}_{\text{lc}}}{(2\pi)^3}\frac{1}{2p_+}c(\bm{p}_{\text{lc}})U^{(\text{in})}_{p,s}(x),
\end{equation}
where the  positive-energy state $U^{(\text{in})}_{p,s}(x)$ with on-shell four-momentum $p^{\mu}$ and spin quantum number $s$ is normalized as
\begin{equation}
\int d^3\bm{x}_{\text{lc}}\bar{U}^{(\text{in})}_{p',s'}(x)\hat{\tilde{n}}U^{(\text{in})}_{p,s}(x)=2p_+\delta_{ss'}(2\pi)^3\delta^3(\bm{p}_{\text{lc}}-\bm{p}'_{\text{lc}}).
\end{equation}
The wave packet is normalized to unity such that the complex coefficients $c(\bm{p}_{\text{lc}})$ must be such that
\begin{equation}
1=\int d^3\bm{x}_{\text{lc}}\bar{\Psi}^{(\text{in})}_s(x)\hat{\tilde{n}}\Psi^{(\text{in})}_s(x)=\int \frac{d^3\bm{p}_{\text{lc}}}{(2\pi)^3}\frac{1}{2p_+}|c(\bm{p}_{\text{lc}})|^2.
\end{equation}
Note that if we consider the wave packet $\Psi^{(\text{in})}_s(x)$ at the light-cone time $T_0$ where the field vanishes, it is
\begin{equation}
\Psi^{(\text{in})}_s(T_0,\bm{x}_{\text{lc}})=\int\frac{d^3\bm{p}_{\text{lc}}}{(2\pi)^3}\frac{1}{2p_+}c(\bm{p}_{\text{lc}})e^{-i(p_-T_0+p_+\phi-\bm{p}_{\perp}\cdot\bm{x}_{\perp})}u_{p,s}.
\end{equation}
We introduce the function
\begin{equation}
\label{f}
f(T_0,\bm{x}_{\text{lc}})=\int\frac{d^3\bm{p}_{\text{lc}}}{(2\pi)^3}\frac{1}{\sqrt{2p_+}}c(\bm{p}_{\text{lc}})e^{-i(p_-T_0+p_+\phi-\bm{p}_{\perp}\cdot\bm{x}_{\perp})},
\end{equation}
which under the approximations introduced below will be interpreted as a sort of initial wave function (the factor $1/\sqrt{2p_+}$ has been included to guarantee that the normalization condition $\int d^3\bm{x}_{\text{lc}}|f(T_0,\bm{x}_{\text{lc}})|^2=1$ is satisfied).

The $S$-matrix amplitude of nonlinear Compton scattering with the incoming electron being in the state $\Psi^{(\text{in})}_s(x)$ and with the final electron (emitted photon) having four-momentum $p^{\prime\,\mu}=(\varepsilon',\bm{p}')$ ($k^{\mu}=(\omega,\bm{k})$) and spin quantum number $s'$ (polarization $l$) is given by
\begin{equation}
S=-ie\int dT\int d^3\bm{x}_{\text{lc}}\,\bar{U}^{(\text{out})}_{p',s'}(x)\frac{\hat{e}^*_{k,l}e^{i(kx)}}{\sqrt{2k_+}}\Psi^{(\text{in})}_s(x),
\end{equation}
where $e^{\mu,*}_{k,l}$ indicates the polarization four-vector of the emitted photon (the quantization volume $\mathcal{V}_0$ introduced in Ref. \cite{Di_Piazza_2021} can be ignored here as we consider an initial wave packet and it does not enter the final probability in any case). The differential emission probability with respect to the photon light-cone three-momentum $\bm{k}_{\text{lc}}$ and averaged (summed) over the initial (final) discrete quantum numbers, is given by
\begin{equation}
\label{dP_NCS}
\begin{split}
&\frac{dP}{d^3\bm{k}_{\text{lc}}}=\frac{e^2}{2}\sum_{s,s',l}\int \frac{d^3\bm{p}'_{\text{lc}}}{(2\pi)^3}\frac{1}{(2\pi)^3}\frac{1}{2k_+}\int\frac{d^3\tilde{\bm{p}}_{\text{lc}}}{(2\pi)^3}c^*(\tilde{\bm{p}}_{\text{lc}})\int\frac{d^3\bm{p}_{\text{lc}}}{(2\pi)^3}c(\bm{p}_{\text{lc}})\int dT dT'\int d^3\bm{x}_{\text{lc}}d^3x'_{\text{lc}}\\
&\quad\times\bar{U}^{(\text{out})}_{p',s'}(x)\hat{e}^*_{k,l}e^{i(kx)}U^{(\text{in})}_{p,s}(x)\bar{U}^{(\text{in})}_{\tilde{p},s}(x')\hat{e}_{k,l}e^{-i(kx')}U^{(\text{out})}_{p',s'}(x').
\end{split}
\end{equation}
Now, we proceed exactly as in Ref. \cite{Di_Piazza_2021} by assuming that the momentum distribution of the incoming electron is well centered around a value $\bm{p}_{0,\text{lc}}$ such that the corresponding energy is the largest dynamical energy in the problem. Analogous conditions and approximations are assumed for the final electron energy. Recalling the concept of formation length \cite{Ter-Mikaelian_b_1972,Baier_b_1998,Baier_2005}, we pass to the average and relative spacetime variables $x_+^{\mu}=(x^{\mu}+x^{\prime\mu})/2$ and $x_-^{\mu}=x^{\mu}-x^{\prime\mu}$, respectively. As we have argued in Ref. \cite{Di_Piazza_2021}, under the above conditions, one can ignore the transverse formation lengths (see also Ref. \cite{Di_Piazza_2021}). Thus, one can set $\bm{x}_{-,\text{lc}}=\bm{0}$ everywhere in Eq. (\ref{dP_NCS}) except in the actions featured in the exponential functions of the electron states, where a first-order expansion on those variables has to be performed:
\begin{equation}
\label{dP_d^3k_WKB}
\begin{split}
&\frac{dP}{d^3\bm{k}_{\text{lc}}}\approx\frac{e^2}{2}\sum_{s,s',l}\int \frac{d^3\bm{p}'_{\text{lc}}}{(2\pi)^6}\int\frac{d^3\tilde{\bm{p}}_{\text{lc}}}{(2\pi)^3}\frac{1}{2\tilde{p}_+}c^*(\tilde{\bm{p}}_{\text{lc}})\int\frac{d^3\bm{p}_{\text{lc}}}{(2\pi)^3}c(\bm{p}_{\text{lc}})\int dT_+ dT_-\int \frac{d^3\bm{x}_{+,\text{lc}}d^3\bm{x}_{-,\text{lc}}}{8p'_+k_+p_+}\\
&\;\;\times\bar{u}^{(\text{out})}_{p',s'}(x_T)\hat{e}^*_{k,l}u^{(\text{in})}_{p,s}(x_T)\bar{u}^{(\text{in})}_{p,s}(x_{T'})\hat{e}_{k,l}u^{(\text{out})}_{p',s'}(x_{T'})\\
&\;\;\times e^{i\big[-S_{p'}^{(\text{out})}(x_T)-\bm{\nabla}_{\perp}S_{p'}^{(\text{out})}(x_T)\cdot\frac{\bm{x}_{-,\perp}}{2}-\partial_{\phi}S_{p'}^{(\text{out})}(x_T)\frac{\phi_-}{2}+(kx)+S_p^{(\text{in})}(x_T)+\bm{\nabla}_{\perp}S_p^{(\text{in})}(x_T)\cdot\frac{\bm{x}_{-,\perp}}{2}+\partial_{\phi}S_p^{(\text{in})}(x_T)\frac{\phi_-}{2}\big]}\\
&\;\;\times e^{i\big[S_{p'}^{(\text{out})}(x_{T'})-\bm{\nabla}_{\perp}S_{p'}^{(\text{out})}(x_{T'})\cdot\frac{\bm{x}_{-,\perp}}{2}-\partial_{\phi}S_{p'}^{(\text{out})}(x_{T'})\frac{\phi_-}{2}-(kx')-S_{\tilde{p}}^{(\text{in})}(x_{T'})+\bm{\nabla}_{\perp}S_{\tilde{p}}^{(\text{in})}(x_{T'})\cdot\frac{\bm{x}_{-,\perp}}{2}+\partial_{\phi}S_{\tilde{p}}^{(\text{in})}(x_{T'})\frac{\phi_-}{2}\big]},
\end{split}
\end{equation}
where $x_T=(T,\bm{x}_{+,\text{lc}})$, $x_{T'}=(T',\bm{x}_{+,\text{lc}})$, and where all the derivatives are assumed to be with respect to the plus variables. 

At this point an important difference arises as compared to Ref. \cite{Di_Piazza_2021} because we have to work out the contribution related to the fact that the incoming electron is described by a wave packet. Notice that the new integrals
\begin{equation}
\int\frac{d^3\tilde{\bm{p}}_{\text{lc}}}{(2\pi)^3}\frac{1}{2\tilde{p}_+}c^*(\tilde{\bm{p}}_{\text{lc}})\int\frac{d^3\bm{p}_{\text{lc}}}{(2\pi)^3}c(\bm{p}_{\text{lc}})
\end{equation}
arise as compared to the corresponding result in Ref. \cite{Di_Piazza_2021}, which instead features an average over the quantization volume. We assume that the wave packet is sufficiently well centered around the momentum $\bm{p}_{0,\text{lc}}$ that the substitutions $\bm{p}_{\text{lc}}\to\bm{p}_{0,\text{lc}}$ and $\tilde{\bm{p}}_{\text{lc}}\to\bm{p}_{0,\text{lc}}$ can be carried out everywhere (including the terms which are already at first order in the minus spacetime variables) except that in the actions $S_p^{(\text{in})}(x_T)$ and $S_{\tilde{p}}^{(\text{in})}(x_{T'})$, where, as we will see, a first-order expansion is necessary:
\begin{align}
S_p^{(\text{in})}(x_T)&\approx S_{p_0}^{(\text{in})}(x_T)+\bm{\nabla}_{\bm{p}_{\perp}}S_{p_0}^{(\text{in})}(x_T)\cdot(\bm{p}_{\perp}-\bm{p}_{0,\perp})+\partial_{p_+}S_{p_0}^{(\text{in})}(x_T)(p_+-p_{0,+}),\\
S_{\tilde{p}}^{(\text{in})}(x_{T'})&\approx S_{p_0}^{(\text{in})}(x_{T'})+\bm{\nabla}_{\bm{p}_{\perp}}S_{p_0}^{(\text{in})}(x_{T'})\cdot(\tilde{\bm{p}}_{\perp}-\bm{p}_{0,\perp})+\partial_{p_+}S_{p_0}^{(\text{in})}(x_{T'})(\tilde{p}_+-p_{0,+}),
\end{align}
where all derivatives are computed at $\bm{p}_{0,\text{lc}}$. In this way, the leading-order terms will provide the same result as in Ref. \cite{Di_Piazza_2021} whereas the corrections result in the additional integrals
\begin{equation}
\label{WP_effect}
\int\frac{d^3\tilde{\bm{p}}_{\text{lc}}}{(2\pi)^3}\frac{1}{2\tilde{p}_+}c^*(\tilde{\bm{p}}_{\text{lc}})\int\frac{d^3\bm{p}_{\text{lc}}}{(2\pi)^3}c(\bm{p}_{\text{lc}})e^{i\bm{\nabla}_{\bm{p}_{\perp}}S_{p_0}^{(\text{in})}(x_+)\cdot(\bm{p}_{\perp}-\tilde{\bm{p}}_{\perp})+i\partial_{p_+}S_{p_0}^{(\text{in})}(x_+)(p_+-\tilde{p}_+)},
\end{equation}
where we have set $T=T'=T_+$ in the derivatives and $x_+=(T_+,\bm{x}_{+,\text{lc}})$. The reason is that in the ultrarelativistic limit the in-action has the following structure \cite{Di_Piazza_2015}
\begin{equation}
S_p^{(\text{in})}(x_T)=-(px_T)-e\int_{-\infty}^Td\tilde{T}A_-(\tilde{T},\bm{x}_{\text{lc}})+O\left(\frac{1}{p_+}\right),
\end{equation}
where for the sake of generality we also accounted for the case that $A_-(x)\neq 0$. Thus, at the leading order in the ultrarelativistic limit the derivatives with respect to the transverse and the plus momenta do not depend on $T$. Now, from the general properties of the actions, which are also reported in Ref. \cite{Di_Piazza_2021}, the derivatives of the action with respect to the momenta are related to the initial conditions for a given trajectory:
\begin{align}
\bm{\nabla}_{\bm{p}_{\perp}}S_{p_0}^{(\text{in})}(x_+)&=\bm{x}_{0,\perp}-\frac{\bm{p}_{0,\perp}}{p_0,+}T_0,\\
\partial_{p_+}S_{p_0}^{(\text{in})}(x_+)&=-\phi_0+\frac{p_{0,-}}{p_{0,+}}T_0.
\end{align}
By using these relations, by noticing that we can approximate $2\tilde{p}_+\approx \sqrt{2\tilde{p}_+}\sqrt{2p_+}$ in the pre-exponential of Eq. (\ref{WP_effect}), and by recalling that in the rest of the probability all quantities are computed at $\bm{p}_{0,\text{lc}}$, we obtain (see also Eq. (\ref{f}))
\begin{equation}
\begin{split}
&\int\frac{d^3\tilde{\bm{p}}_{\text{lc}}}{(2\pi)^3}\frac{1}{2\tilde{p}_+}c^*(\tilde{\bm{p}}_{\text{lc}})\int\frac{d^3\bm{p}_{\text{lc}}}{(2\pi)^3}c(\bm{p}_{\text{lc}})e^{i\bm{\nabla}_{\bm{p}_{\perp}}S_{p_0}^{(\text{in})}(x_T)\cdot(\bm{p}_{\perp}-\tilde{\bm{p}}_{\perp})+i\partial_{p_+}S_{p_0}^{(\text{in})}(x_T)(p_+-\tilde{p}_+)}\\
&\quad=\int\frac{d^3\tilde{\bm{p}}_{\text{lc}}}{(2\pi)^3}\frac{1}{\sqrt{2\tilde{p}_+}}c^*(\tilde{\bm{p}}_{\text{lc}})e^{i(\tilde{p}_-T_0+\tilde{p}_+\phi_0-\tilde{\bm{p}}_{\perp}\cdot\bm{x}_{0,\perp})}\int\frac{d^3\bm{p}_{\text{lc}}}{(2\pi)^3}\frac{1}{\sqrt{2p_+}}c(\bm{p}_{\text{lc}})e^{-i(p_-T_0+p_+\phi_0-\bm{p}_{\perp}\cdot\bm{x}_{0,\perp})}\\
&\quad=|f(T_0,\bm{x}_{0,\text{lc}})|^2.
\end{split}
\end{equation}

Before concluding the derivation, we would like to report an additional step in the derivation presented in Ref. \cite{Di_Piazza_2021}, which is also related to the considerations on the structure of the action in the ultrarelativistic limit discussed below Eq. (\ref{WP_effect}). The momenta involved in the conservation laws arising from the integrals over $\bm{x}_{-,\perp}$ and $\phi_-$ in Eq. (\ref{dP_d^3k_WKB}) are computed at the different times $T$ and $T'$. As above, however, it is possible to approximate them as $T_+$ at the leading order in the electron energy. To see this we also report the structure of the out-action in the ultrarelativistic limit \cite{Di_Piazza_2015}
\begin{equation}
S_p^{(\text{out})}(x_T)=-(px_T)-e\int_{\infty}^Td\tilde{T}A_-(\tilde{T},\bm{x}_{\text{lc}})+O\left(\frac{1}{p_+}\right),
\end{equation}
again in the case where $A_-(x)\neq 0$. Since the argument for the transverse and the plus momenta is the same, we only work out the former case in detail. After approximating $\tilde{\bm{p}}_{\text{lc}}$ with $\bm{p}_{\text{lc}}$ in the terms giving rise to the conserving delta functions, we have that that conservation law reads
\begin{equation}
\frac{-\bm{\nabla}_{\perp}S_{p'}^{(\text{out})}(x_T)-\bm{\nabla}_{\perp}S_{p'}^{(\text{out})}(x_{T'})}{2}-\bm{k}_{\perp}+\frac{\bm{\nabla}_{\perp}S_p^{(\text{in})}(x_T)+\bm{\nabla}_{\perp}S_p^{(\text{in})}(x_{T'})}{2}=\bm{0}.
\end{equation}
By using the expressions of the in- and out-actions, one sees that one can combine the integrals in the conservation law and that, apart from negligible terms scaling as $1/p_+$, one obtains
\begin{equation}
\begin{split}
\bm{0}&=-\bm{p}'_{\perp}-e\bm{\nabla}_{\perp}\int_{-\infty}^{\infty}d\tilde{T}A_-(\tilde{T},\bm{x}_{\text{lc}})-\bm{k}_{\perp}+\bm{p}_{\perp}\\
&=-\bm{p}'_{\perp}+e\bm{\nabla}_{\perp}\int_{\infty}^{T_+}d\tilde{T}A_-(\tilde{T},\bm{x}_{\text{lc}})-\bm{k}_{\perp}+\bm{p}_{\perp}-e\bm{\nabla}_{\perp}\int_{-\infty}^{T_+}d\tilde{T}A_-(\tilde{T},\bm{x}_{\text{lc}})\\
&\approx-\bm{\nabla}_{\perp}S_{p'}^{(\text{out})}(x_{T_+})-\bm{k}_{\perp}+\bm{\nabla}_{\perp}S_p^{(\text{in})}(x_{T_+}),
\end{split}
\end{equation}
as reported in Ref. \cite{Di_Piazza_2021}.

At this point one can follow the same steps as in Ref. \cite{Di_Piazza_2021}, where it has been demonstrated that Baier's formula is recovered even for fixed spin and polarization quantum numbers of the involved particles. The only difference with the results in Ref. \cite{Di_Piazza_2021} is that we considered here an electron initially described by a wave packet, which is taken into account by means of the substitution 
\begin{equation}
\lim_{\mathcal{V}_0\to\infty}\int_{\mathcal{V}_0} \frac{d^3\bm{x}_{0,\text{lc}}}{\mathcal{V}_0} \to \int d^3\bm{x}_{0,\text{lc}}|f(T_0,\bm{x}_{0,\text{lc}})|^2.
\end{equation}
This result is indeed expected, once one accounts for the discussed physical meaning of the function $f(T,\bm{x}_{\text{lc}})$. Finally, we recall that in Ref. \cite{Wistisen_2014} it has been shown that Baier's formula can be conveniently written in a similar manner as the classical formula of radiation, which features the electron's acceleration. Accounting for the substitution rule mentioned above, we arrive at the result
\begin{equation}
\label{dP_d^3k_Baier}
\begin{split}
\frac{dP}{d^3\bm{k}_{\text{lc}}}&=\int d^3\bm{x}_{0,\text{lc}}|f(T_0,\bm{x}_{0,\text{lc}})|^2\frac{\alpha}{4\pi^2k_+^3}\gig\{\frac{p_+^2+p_+^{\prime\,2}}{2p_+^2}\\
&\quad\times\left|\int dT\frac{\bm{n}_k\times\{[\bm{n}_k-\bm{V}^{(\text{in})}_e(T)]\times\dot{\bm{V}}^{(\text{in})}_e(T)\}}{[1-\bm{n}_k\cdot\bm{V}^{(\text{in})}_e(T)]^2}e^{i\frac{k_+p_+}{p'_+}\int_{T_0}^Td\tilde{T}[1-\bm{n}_k\cdot\bm{V}^{(\text{in})}_e(\tilde{T})]}\right|^2\\
&\left.\quad+\frac{k_+^2m^2}{2p_+^4}\left|\int dT\frac{\bm{n}_k\cdot\dot{\bm{V}}^{(\text{in})}_e(T)}{[1-\bm{n}_k\cdot\bm{V}^{(\text{in})}_e(T)]^2}e^{i\frac{k_+p_+}{p'_+}\int_{T_0}^Td\tilde{T}[1-\bm{n}_k\cdot\bm{V}^{(\text{in})}_e(\tilde{T})]}\right|^2\right\},
\end{split}
\end{equation}
where we have introduced $\bm{n}_k=\bm{k}/\omega$ and the electron velocity $\bm{V}^{(\text{in})}_e(T)=\bm{\Pi}^{(\text{in})}_e(T)/\Pi^{0\,(\text{in})}_e(T)$, which were not defined in Ref. \cite{Di_Piazza_2021}.

Finally, the result in the text (see Eq. (\ref{dP_d^3k_Baier_2})) is obtained by means of the substitutions $T\to t$ (and then $\bm{x}_{0,\text{lc}}\to \bm{x}_0$), $p_+\to\varepsilon$, $k_+\to\omega$, $p'_+\to\varepsilon'$, which are all consistent in the leading-order ultrarelativistic limit, as well as $\bm{n}_k\to \bm{n}$ and $\bm{V}^{(\text{in})}_e(t)\to \bm{v}(t)$ (and then by the obvious relation $dP/d\omega d\Omega=\omega^2dP/d^3\bm{k}$).


%\bibliography{arXiv,Books,Reviews,Papers_Nuclei,Papers_Radiation,Papers_RR,Papers_PP_and_Cascades,Papers_VPE,Papers_Crystal,Papers_Various,Papers_Laser_Plasma_QED,Papers_XFEL,Homepages,Papers_Gravity}

\begin{thebibliography}{0}%
\makeatletter
\providecommand \@ifxundefined [1]{%
 \@ifx{#1\undefined}
}%
\providecommand \@ifnum [1]{%
 \ifnum #1\expandafter \@firstoftwo
 \else \expandafter \@secondoftwo
 \fi
}%
\providecommand \@ifx [1]{%
 \ifx #1\expandafter \@firstoftwo
 \else \expandafter \@secondoftwo
 \fi
}%
\providecommand \natexlab [1]{#1}%
\providecommand \enquote  [1]{``#1''}%
\providecommand \bibnamefont  [1]{#1}%
\providecommand \bibfnamefont [1]{#1}%
\providecommand \citenamefont [1]{#1}%
\providecommand \href@noop [0]{\@secondoftwo}%
\providecommand \href [0]{\begingroup \@sanitize@url \@href}%
\providecommand \@href[1]{\@@startlink{#1}\@@href}%
\providecommand \@@href[1]{\endgroup#1\@@endlink}%
\providecommand \@sanitize@url [0]{\catcode `\\12\catcode `\$12\catcode
  `\&12\catcode `\#12\catcode `\^12\catcode `\_12\catcode `\%12\relax}%
\providecommand \@@startlink[1]{}%
\providecommand \@@endlink[0]{}%
\providecommand \url  [0]{\begingroup\@sanitize@url \@url }%
\providecommand \@url [1]{\endgroup\@href {#1}{\urlprefix }}%
\providecommand \urlprefix  [0]{URL }%
\providecommand \Eprint [0]{\href }%
\providecommand \doibase [0]{https://doi.org/}%
\providecommand \selectlanguage [0]{\@gobble}%
\providecommand \bibinfo  [0]{\@secondoftwo}%
\providecommand \bibfield  [0]{\@secondoftwo}%
\providecommand \translation [1]{[#1]}%
\providecommand \BibitemOpen [0]{}%
\providecommand \bibitemStop [0]{}%
\providecommand \bibitemNoStop [0]{.\EOS\space}%
\providecommand \EOS [0]{\spacefactor3000\relax}%
\providecommand \BibitemShut  [1]{\csname bibitem#1\endcsname}%
\let\auto@bib@innerbib\@empty
%</preamble>
\end{thebibliography}%


\begin{thebibliography}{52}%
\makeatletter
\providecommand \@ifxundefined [1]{%
 \@ifx{#1\undefined}
}%
\providecommand \@ifnum [1]{%
 \ifnum #1\expandafter \@firstoftwo
 \else \expandafter \@secondoftwo
 \fi
}%
\providecommand \@ifx [1]{%
 \ifx #1\expandafter \@firstoftwo
 \else \expandafter \@secondoftwo
 \fi
}%
\providecommand \natexlab [1]{#1}%
\providecommand \enquote  [1]{``#1''}%
\providecommand \bibnamefont  [1]{#1}%
\providecommand \bibfnamefont [1]{#1}%
\providecommand \citenamefont [1]{#1}%
\providecommand \href@noop [0]{\@secondoftwo}%
\providecommand \href [0]{\begingroup \@sanitize@url \@href}%
\providecommand \@href[1]{\@@startlink{#1}\@@href}%
\providecommand \@@href[1]{\endgroup#1\@@endlink}%
\providecommand \@sanitize@url [0]{\catcode `\\12\catcode `\$12\catcode
  `\&12\catcode `\#12\catcode `\^12\catcode `\_12\catcode `\%12\relax}%
\providecommand \@@startlink[1]{}%
\providecommand \@@endlink[0]{}%
\providecommand \url  [0]{\begingroup\@sanitize@url \@url }%
\providecommand \@url [1]{\endgroup\@href {#1}{\urlprefix }}%
\providecommand \urlprefix  [0]{URL }%
\providecommand \Eprint [0]{\href }%
\providecommand \doibase [0]{https://doi.org/}%
\providecommand \selectlanguage [0]{\@gobble}%
\providecommand \bibinfo  [0]{\@secondoftwo}%
\providecommand \bibfield  [0]{\@secondoftwo}%
\providecommand \translation [1]{[#1]}%
\providecommand \BibitemOpen [0]{}%
\providecommand \bibitemStop [0]{}%
\providecommand \bibitemNoStop [0]{.\EOS\space}%
\providecommand \EOS [0]{\spacefactor3000\relax}%
\providecommand \BibitemShut  [1]{\csname bibitem#1\endcsname}%
\let\auto@bib@innerbib\@empty
%</preamble>
\bibitem [{\citenamefont {Jackson}(1975)}]{Jackson_b_1975}%
  \BibitemOpen
  \bibfield  {author} {\bibinfo {author} {\bibfnamefont {J.~D.}\ \bibnamefont
  {Jackson}},\ }\href@noop {} {\emph {\bibinfo {title} {Classical
  Electrodynamics}}}\ (\bibinfo  {publisher} {John Wiley \& Sons, New York},\
  \bibinfo {year} {1975})\BibitemShut {NoStop}%
\bibitem [{\citenamefont {Landau}\ and\ \citenamefont
  {Lifshitz}(1975)}]{Landau_b_2_1975}%
  \BibitemOpen
  \bibfield  {author} {\bibinfo {author} {\bibfnamefont {L.~D.}\ \bibnamefont
  {Landau}}\ and\ \bibinfo {author} {\bibfnamefont {E.~M.}\ \bibnamefont
  {Lifshitz}},\ }\href@noop {} {\emph {\bibinfo {title} {The Classical Theory
  of Fields}}}\ (\bibinfo  {publisher} {Elsevier, Oxford},\ \bibinfo {year}
  {1975})\BibitemShut {NoStop}%
\bibitem [{\citenamefont {Berestetskii}\ \emph {et~al.}(1982)\citenamefont
  {Berestetskii}, \citenamefont {Lifshitz},\ and\ \citenamefont
  {Pitaevskii}}]{Landau_b_4_1982}%
  \BibitemOpen
  \bibfield  {author} {\bibinfo {author} {\bibfnamefont {V.~B.}\ \bibnamefont
  {Berestetskii}}, \bibinfo {author} {\bibfnamefont {E.~M.}\ \bibnamefont
  {Lifshitz}},\ and\ \bibinfo {author} {\bibfnamefont {L.~P.}\ \bibnamefont
  {Pitaevskii}},\ }\href@noop {} {\emph {\bibinfo {title} {Quantum
  Electrodynamics}}}\ (\bibinfo  {publisher} {Elsevier Butterworth-Heinemann,
  Oxford},\ \bibinfo {year} {1982})\BibitemShut {NoStop}%
\bibitem [{\citenamefont {Peskin}\ and\ \citenamefont
  {Schroeder}(1995)}]{Peskin_b_1995}%
  \BibitemOpen
  \bibfield  {author} {\bibinfo {author} {\bibfnamefont {M.~E.}\ \bibnamefont
  {Peskin}}\ and\ \bibinfo {author} {\bibfnamefont {D.~V.}\ \bibnamefont
  {Schroeder}},\ }\href@noop {} {\emph {\bibinfo {title} {An Introduction to
  Quantum Field Theory}}}\ (\bibinfo  {publisher} {Westview Press, Boulder},\
  \bibinfo {year} {1995})\BibitemShut {NoStop}%
\bibitem [{\citenamefont {Dirac}(1938)}]{Dirac_1938}%
  \BibitemOpen
  \bibfield  {author} {\bibinfo {author} {\bibfnamefont {P.~A.~M.}\
  \bibnamefont {Dirac}},\ }\href@noop {} {\bibfield  {journal} {\bibinfo
  {journal} {Proc. R. Soc. London, Ser. A}\ }\textbf {\bibinfo {volume}
  {167}},\ \bibinfo {pages} {148} (\bibinfo {year} {1938})}\BibitemShut
  {NoStop}%
\bibitem [{\citenamefont {Abraham}(1905)}]{Abraham_b_1905}%
  \BibitemOpen
  \bibfield  {author} {\bibinfo {author} {\bibfnamefont {M.}~\bibnamefont
  {Abraham}},\ }\href@noop {} {\emph {\bibinfo {title} {Theorie der
  Elektrizit{\"a}t}}}\ (\bibinfo  {publisher} {Teubner, Leipzig},\ \bibinfo
  {year} {1905})\BibitemShut {NoStop}%
\bibitem [{\citenamefont {Di~Piazza}\ \emph {et~al.}(2012)\citenamefont
  {Di~Piazza}, \citenamefont {M\"{u}ller}, \citenamefont {Hatsagortsyan},\ and\
  \citenamefont {Keitel}}]{Di_Piazza_2012}%
  \BibitemOpen
  \bibfield  {author} {\bibinfo {author} {\bibfnamefont {A.}~\bibnamefont
  {Di~Piazza}}, \bibinfo {author} {\bibfnamefont {C.}~\bibnamefont
  {M\"{u}ller}}, \bibinfo {author} {\bibfnamefont {K.~Z.}\ \bibnamefont
  {Hatsagortsyan}},\ and\ \bibinfo {author} {\bibfnamefont {C.~H.}\
  \bibnamefont {Keitel}},\ }\href@noop {} {\bibfield  {journal} {\bibinfo
  {journal} {Rev. Mod. Phys.}\ }\textbf {\bibinfo {volume} {84}},\ \bibinfo
  {pages} {1177} (\bibinfo {year} {2012})}\BibitemShut {NoStop}%
\bibitem [{\citenamefont {Burton}\ and\ \citenamefont
  {Noble}(2014)}]{Burton_2014}%
  \BibitemOpen
  \bibfield  {author} {\bibinfo {author} {\bibfnamefont {D.~A.}\ \bibnamefont
  {Burton}}\ and\ \bibinfo {author} {\bibfnamefont {A.}~\bibnamefont {Noble}},\
  }\href@noop {} {\bibfield  {journal} {\bibinfo  {journal} {Contemp. Phys.}\
  }\textbf {\bibinfo {volume} {55}},\ \bibinfo {pages} {110} (\bibinfo {year}
  {2014})}\BibitemShut {NoStop}%
\bibitem [{\citenamefont {Gonoskov}\ \emph {et~al.}(2022)\citenamefont
  {Gonoskov}, \citenamefont {Blackburn}, \citenamefont {Marklund},\ and\
  \citenamefont {Bulanov}}]{Gonoskov_2022}%
  \BibitemOpen
  \bibfield  {author} {\bibinfo {author} {\bibfnamefont {A.}~\bibnamefont
  {Gonoskov}}, \bibinfo {author} {\bibfnamefont {T.~G.}\ \bibnamefont
  {Blackburn}}, \bibinfo {author} {\bibfnamefont {M.}~\bibnamefont
  {Marklund}},\ and\ \bibinfo {author} {\bibfnamefont {S.~S.}\ \bibnamefont
  {Bulanov}},\ }\href@noop {} {\bibfield  {journal} {\bibinfo  {journal} {Rev.
  Mod. Phys.}\ }\textbf {\bibinfo {volume} {94}},\ \bibinfo {pages} {045001}
  (\bibinfo {year} {2022})}\BibitemShut {NoStop}%
\bibitem [{\citenamefont {Fedotov}\ \emph {et~al.}(2023)\citenamefont
  {Fedotov}, \citenamefont {Ilderton}, \citenamefont {Karbstein}, \citenamefont
  {King}, \citenamefont {Seipt}, \citenamefont {Taya},\ and\ \citenamefont
  {Torgrimsson}}]{Fedotov_2023}%
  \BibitemOpen
  \bibfield  {author} {\bibinfo {author} {\bibfnamefont {A.}~\bibnamefont
  {Fedotov}}, \bibinfo {author} {\bibfnamefont {A.}~\bibnamefont {Ilderton}},
  \bibinfo {author} {\bibfnamefont {F.}~\bibnamefont {Karbstein}}, \bibinfo
  {author} {\bibfnamefont {B.}~\bibnamefont {King}}, \bibinfo {author}
  {\bibfnamefont {D.}~\bibnamefont {Seipt}}, \bibinfo {author} {\bibfnamefont
  {H.}~\bibnamefont {Taya}},\ and\ \bibinfo {author} {\bibfnamefont
  {G.}~\bibnamefont {Torgrimsson}},\ }\href@noop {} {\bibfield  {journal}
  {\bibinfo  {journal} {Phys. Rep.}\ }\textbf {\bibinfo {volume} {1010}},\
  \bibinfo {pages} {1} (\bibinfo {year} {2023})}\BibitemShut {NoStop}%
\bibitem [{\citenamefont {Cole}\ \emph {et~al.}(2018)\citenamefont {Cole},
  \citenamefont {Behm}, \citenamefont {Gerstmayr}, \citenamefont {Blackburn},
  \citenamefont {Wood}, \citenamefont {Baird}, \citenamefont {Duff},
  \citenamefont {Harvey}, \citenamefont {Ilderton}, \citenamefont {Joglekar},
  \citenamefont {Krushelnick}, \citenamefont {Kuschel}, \citenamefont
  {Marklund}, \citenamefont {McKenna}, \citenamefont {Murphy}, \citenamefont
  {Poder}, \citenamefont {Ridgers}, \citenamefont {Samarin}, \citenamefont
  {Sarri}, \citenamefont {Symes}, \citenamefont {Thomas}, \citenamefont
  {Warwick}, \citenamefont {Zepf}, \citenamefont {Najmudin},\ and\
  \citenamefont {Mangles}}]{Cole_2018}%
  \BibitemOpen
  \bibfield  {author} {\bibinfo {author} {\bibfnamefont {J.~M.}\ \bibnamefont
  {Cole}}, \bibinfo {author} {\bibfnamefont {K.~T.}\ \bibnamefont {Behm}},
  \bibinfo {author} {\bibfnamefont {E.}~\bibnamefont {Gerstmayr}}, \bibinfo
  {author} {\bibfnamefont {T.~G.}\ \bibnamefont {Blackburn}}, \bibinfo {author}
  {\bibfnamefont {J.~C.}\ \bibnamefont {Wood}}, \bibinfo {author}
  {\bibfnamefont {C.~D.}\ \bibnamefont {Baird}}, \bibinfo {author}
  {\bibfnamefont {M.~J.}\ \bibnamefont {Duff}}, \bibinfo {author}
  {\bibfnamefont {C.}~\bibnamefont {Harvey}}, \bibinfo {author} {\bibfnamefont
  {A.}~\bibnamefont {Ilderton}}, \bibinfo {author} {\bibfnamefont {A.~S.}\
  \bibnamefont {Joglekar}}, \bibinfo {author} {\bibfnamefont {K.}~\bibnamefont
  {Krushelnick}}, \bibinfo {author} {\bibfnamefont {S.}~\bibnamefont
  {Kuschel}}, \bibinfo {author} {\bibfnamefont {M.}~\bibnamefont {Marklund}},
  \bibinfo {author} {\bibfnamefont {P.}~\bibnamefont {McKenna}}, \bibinfo
  {author} {\bibfnamefont {C.~D.}\ \bibnamefont {Murphy}}, \bibinfo {author}
  {\bibfnamefont {K.}~\bibnamefont {Poder}}, \bibinfo {author} {\bibfnamefont
  {C.~P.}\ \bibnamefont {Ridgers}}, \bibinfo {author} {\bibfnamefont {G.~M.}\
  \bibnamefont {Samarin}}, \bibinfo {author} {\bibfnamefont {G.}~\bibnamefont
  {Sarri}}, \bibinfo {author} {\bibfnamefont {D.~R.}\ \bibnamefont {Symes}},
  \bibinfo {author} {\bibfnamefont {A.~G.~R.}\ \bibnamefont {Thomas}}, \bibinfo
  {author} {\bibfnamefont {J.}~\bibnamefont {Warwick}}, \bibinfo {author}
  {\bibfnamefont {M.}~\bibnamefont {Zepf}}, \bibinfo {author} {\bibfnamefont
  {Z.}~\bibnamefont {Najmudin}},\ and\ \bibinfo {author} {\bibfnamefont
  {S.~P.~D.}\ \bibnamefont {Mangles}},\ }\href@noop {} {\bibfield  {journal}
  {\bibinfo  {journal} {Phys. Rev. X}\ }\textbf {\bibinfo {volume} {8}},\
  \bibinfo {pages} {011020} (\bibinfo {year} {2018})}\BibitemShut {NoStop}%
\bibitem [{\citenamefont {Poder}\ \emph {et~al.}(2018)\citenamefont {Poder},
  \citenamefont {Tamburini}, \citenamefont {Sarri}, \citenamefont {Di~Piazza},
  \citenamefont {Kuschel}, \citenamefont {Baird}, \citenamefont {Behm},
  \citenamefont {Bohlen}, \citenamefont {Cole}, \citenamefont {Corvan},
  \citenamefont {Duff}, \citenamefont {Gerstmayr}, \citenamefont {Keitel},
  \citenamefont {Krushelnick}, \citenamefont {Mangles}, \citenamefont
  {McKenna}, \citenamefont {Murphy}, \citenamefont {Najmudin}, \citenamefont
  {Ridgers}, \citenamefont {Samarin}, \citenamefont {Symes}, \citenamefont
  {Thomas}, \citenamefont {Warwick},\ and\ \citenamefont {Zepf}}]{Poder_2018}%
  \BibitemOpen
  \bibfield  {author} {\bibinfo {author} {\bibfnamefont {K.}~\bibnamefont
  {Poder}}, \bibinfo {author} {\bibfnamefont {M.}~\bibnamefont {Tamburini}},
  \bibinfo {author} {\bibfnamefont {G.}~\bibnamefont {Sarri}}, \bibinfo
  {author} {\bibfnamefont {A.}~\bibnamefont {Di~Piazza}}, \bibinfo {author}
  {\bibfnamefont {S.}~\bibnamefont {Kuschel}}, \bibinfo {author} {\bibfnamefont
  {C.~D.}\ \bibnamefont {Baird}}, \bibinfo {author} {\bibfnamefont
  {K.}~\bibnamefont {Behm}}, \bibinfo {author} {\bibfnamefont {S.}~\bibnamefont
  {Bohlen}}, \bibinfo {author} {\bibfnamefont {J.~M.}\ \bibnamefont {Cole}},
  \bibinfo {author} {\bibfnamefont {D.~J.}\ \bibnamefont {Corvan}}, \bibinfo
  {author} {\bibfnamefont {M.}~\bibnamefont {Duff}}, \bibinfo {author}
  {\bibfnamefont {E.}~\bibnamefont {Gerstmayr}}, \bibinfo {author}
  {\bibfnamefont {C.~H.}\ \bibnamefont {Keitel}}, \bibinfo {author}
  {\bibfnamefont {K.}~\bibnamefont {Krushelnick}}, \bibinfo {author}
  {\bibfnamefont {S.~P.~D.}\ \bibnamefont {Mangles}}, \bibinfo {author}
  {\bibfnamefont {P.}~\bibnamefont {McKenna}}, \bibinfo {author} {\bibfnamefont
  {C.~D.}\ \bibnamefont {Murphy}}, \bibinfo {author} {\bibfnamefont
  {Z.}~\bibnamefont {Najmudin}}, \bibinfo {author} {\bibfnamefont {C.~P.}\
  \bibnamefont {Ridgers}}, \bibinfo {author} {\bibfnamefont {G.~M.}\
  \bibnamefont {Samarin}}, \bibinfo {author} {\bibfnamefont {D.~R.}\
  \bibnamefont {Symes}}, \bibinfo {author} {\bibfnamefont {A.~G.~R.}\
  \bibnamefont {Thomas}}, \bibinfo {author} {\bibfnamefont {J.}~\bibnamefont
  {Warwick}},\ and\ \bibinfo {author} {\bibfnamefont {M.}~\bibnamefont
  {Zepf}},\ }\href@noop {} {\bibfield  {journal} {\bibinfo  {journal} {Phys.
  Rev. X}\ }\textbf {\bibinfo {volume} {8}},\ \bibinfo {pages} {031004}
  (\bibinfo {year} {2018})}\BibitemShut {NoStop}%
\bibitem [{\citenamefont {Wistisen}\ \emph {et~al.}(2018)\citenamefont
  {Wistisen}, \citenamefont {Di~Piazza}, \citenamefont {Knudsen},\ and\
  \citenamefont {Uggerh{\o}j}}]{Wistisen_2018}%
  \BibitemOpen
  \bibfield  {author} {\bibinfo {author} {\bibfnamefont {T.~N.}\ \bibnamefont
  {Wistisen}}, \bibinfo {author} {\bibfnamefont {A.}~\bibnamefont {Di~Piazza}},
  \bibinfo {author} {\bibfnamefont {H.~V.}\ \bibnamefont {Knudsen}},\ and\
  \bibinfo {author} {\bibfnamefont {U.~I.}\ \bibnamefont {Uggerh{\o}j}},\
  }\href@noop {} {\bibfield  {journal} {\bibinfo  {journal} {Nat. Commun.}\
  }\textbf {\bibinfo {volume} {9}},\ \bibinfo {pages} {795} (\bibinfo {year}
  {2018})}\BibitemShut {NoStop}%
\bibitem [{\citenamefont {Los}\ \emph {et~al.}(2026)\citenamefont {Los},
  \citenamefont {Gerstmayr}, \citenamefont {Arran}, \citenamefont {Streeter},
  \citenamefont {Colgan}, \citenamefont {Cobo}, \citenamefont {Kettle},
  \citenamefont {Blackburn}, \citenamefont {Bourgeois}, \citenamefont {Calvin},
  \citenamefont {Cardarelli}, \citenamefont {Cavanagh}, \citenamefont {Dann},
  \citenamefont {Di~Piazza}, \citenamefont {Fitzgarrald}, \citenamefont
  {Ilderton}, \citenamefont {Keitel}, \citenamefont {Marklund}, \citenamefont
  {McKenna}, \citenamefont {Murphy}, \citenamefont {Najmudin}, \citenamefont
  {Parsons}, \citenamefont {Rajeev}, \citenamefont {Symes}, \citenamefont
  {Tamburini}, \citenamefont {Thomas}, \citenamefont {Wood}, \citenamefont
  {Zepf}, \citenamefont {Sarri}, \citenamefont {Ridgers},\ and\ \citenamefont
  {Mangles}}]{Los2026}%
  \BibitemOpen
  \bibfield  {author} {\bibinfo {author} {\bibfnamefont {E.~E.}\ \bibnamefont
  {Los}}, \bibinfo {author} {\bibfnamefont {E.}~\bibnamefont {Gerstmayr}},
  \bibinfo {author} {\bibfnamefont {C.}~\bibnamefont {Arran}}, \bibinfo
  {author} {\bibfnamefont {M.~J.~V.}\ \bibnamefont {Streeter}}, \bibinfo
  {author} {\bibfnamefont {C.}~\bibnamefont {Colgan}}, \bibinfo {author}
  {\bibfnamefont {C.~C.}\ \bibnamefont {Cobo}}, \bibinfo {author}
  {\bibfnamefont {B.}~\bibnamefont {Kettle}}, \bibinfo {author} {\bibfnamefont
  {T.~G.}\ \bibnamefont {Blackburn}}, \bibinfo {author} {\bibfnamefont
  {N.}~\bibnamefont {Bourgeois}}, \bibinfo {author} {\bibfnamefont
  {L.}~\bibnamefont {Calvin}}, \bibinfo {author} {\bibfnamefont
  {J.}~\bibnamefont {Cardarelli}}, \bibinfo {author} {\bibfnamefont
  {N.}~\bibnamefont {Cavanagh}}, \bibinfo {author} {\bibfnamefont {S.~J.~D.}\
  \bibnamefont {Dann}}, \bibinfo {author} {\bibfnamefont {A.}~\bibnamefont
  {Di~Piazza}}, \bibinfo {author} {\bibfnamefont {R.}~\bibnamefont
  {Fitzgarrald}}, \bibinfo {author} {\bibfnamefont {A.}~\bibnamefont
  {Ilderton}}, \bibinfo {author} {\bibfnamefont {C.~H.}\ \bibnamefont
  {Keitel}}, \bibinfo {author} {\bibfnamefont {M.}~\bibnamefont {Marklund}},
  \bibinfo {author} {\bibfnamefont {P.}~\bibnamefont {McKenna}}, \bibinfo
  {author} {\bibfnamefont {C.~D.}\ \bibnamefont {Murphy}}, \bibinfo {author}
  {\bibfnamefont {Z.}~\bibnamefont {Najmudin}}, \bibinfo {author}
  {\bibfnamefont {P.}~\bibnamefont {Parsons}}, \bibinfo {author} {\bibfnamefont
  {P.~P.}\ \bibnamefont {Rajeev}}, \bibinfo {author} {\bibfnamefont {D.~R.}\
  \bibnamefont {Symes}}, \bibinfo {author} {\bibfnamefont {M.}~\bibnamefont
  {Tamburini}}, \bibinfo {author} {\bibfnamefont {A.~G.~R.}\ \bibnamefont
  {Thomas}}, \bibinfo {author} {\bibfnamefont {J.~C.}\ \bibnamefont {Wood}},
  \bibinfo {author} {\bibfnamefont {M.}~\bibnamefont {Zepf}}, \bibinfo {author}
  {\bibfnamefont {G.}~\bibnamefont {Sarri}}, \bibinfo {author} {\bibfnamefont
  {C.~P.}\ \bibnamefont {Ridgers}},\ and\ \bibinfo {author} {\bibfnamefont
  {S.~P.~D.}\ \bibnamefont {Mangles}},\ }\href@noop {} {\bibfield  {journal}
  {\bibinfo  {journal} {Nature Commun.}\ }\textbf {\bibinfo {volume} {17}},\
  \bibinfo {pages} {1157} (\bibinfo {year} {2026})}\BibitemShut {NoStop}%
\bibitem [{\citenamefont {Di~Piazza}(2018)}]{Di_Piazza_2018_b}%
  \BibitemOpen
  \bibfield  {author} {\bibinfo {author} {\bibfnamefont {A.}~\bibnamefont
  {Di~Piazza}},\ }\href@noop {} {\bibfield  {journal} {\bibinfo  {journal}
  {Phys. Lett. B}\ }\textbf {\bibinfo {volume} {782}},\ \bibinfo {pages} {559}
  (\bibinfo {year} {2018})}\BibitemShut {NoStop}%
\bibitem [{\citenamefont {Bloch}\ and\ \citenamefont
  {Nordsieck}(1937)}]{Bloch_1937}%
  \BibitemOpen
  \bibfield  {author} {\bibinfo {author} {\bibfnamefont {F.}~\bibnamefont
  {Bloch}}\ and\ \bibinfo {author} {\bibfnamefont {A.}~\bibnamefont
  {Nordsieck}},\ }\href@noop {} {\bibfield  {journal} {\bibinfo  {journal}
  {Phys. Rev.}\ }\textbf {\bibinfo {volume} {52}},\ \bibinfo {pages} {54}
  (\bibinfo {year} {1937})}\BibitemShut {NoStop}%
\bibitem [{\citenamefont {Yennie}\ \emph {et~al.}(1961)\citenamefont {Yennie},
  \citenamefont {Frautschi},\ and\ \citenamefont {Suura}}]{Yennie_1961}%
  \BibitemOpen
  \bibfield  {author} {\bibinfo {author} {\bibfnamefont {D.}~\bibnamefont
  {Yennie}}, \bibinfo {author} {\bibfnamefont {S.}~\bibnamefont {Frautschi}},\
  and\ \bibinfo {author} {\bibfnamefont {H.}~\bibnamefont {Suura}},\
  }\href@noop {} {\bibfield  {journal} {\bibinfo  {journal} {Ann. Phys.}\
  }\textbf {\bibinfo {volume} {13}},\ \bibinfo {pages} {379} (\bibinfo {year}
  {1961})}\BibitemShut {NoStop}%
\bibitem [{\citenamefont {Weinberg}(1965)}]{Weinberg:1965nx}%
  \BibitemOpen
  \bibfield  {author} {\bibinfo {author} {\bibfnamefont {S.}~\bibnamefont
  {Weinberg}},\ }\href {https://doi.org/10.1103/PhysRev.140.B516} {\bibfield
  {journal} {\bibinfo  {journal} {Phys. Rev.}\ }\textbf {\bibinfo {volume}
  {140}},\ \bibinfo {pages} {B516} (\bibinfo {year} {1965})}\BibitemShut
  {NoStop}%
\bibitem [{\citenamefont {Agarwal}\ \emph {et~al.}(2023)\citenamefont
  {Agarwal}, \citenamefont {Magnea}, \citenamefont {Signorile-Signorile},\ and\
  \citenamefont {Tripathi}}]{Agarwal_2021}%
  \BibitemOpen
  \bibfield  {author} {\bibinfo {author} {\bibfnamefont {N.}~\bibnamefont
  {Agarwal}}, \bibinfo {author} {\bibfnamefont {L.}~\bibnamefont {Magnea}},
  \bibinfo {author} {\bibfnamefont {C.}~\bibnamefont {Signorile-Signorile}},\
  and\ \bibinfo {author} {\bibfnamefont {A.}~\bibnamefont {Tripathi}},\ }\href
  {https://doi.org/10.1016/j.physrep.2022.10.001} {\bibfield  {journal}
  {\bibinfo  {journal} {Phys. Rept.}\ }\textbf {\bibinfo {volume} {994}},\
  \bibinfo {pages} {1} (\bibinfo {year} {2023})},\ \Eprint
  {https://arxiv.org/abs/2112.07099} {arXiv:2112.07099 [hep-ph]} \BibitemShut
  {NoStop}%
\bibitem [{\citenamefont {Itzykson}\ and\ \citenamefont
  {Zuber}(1980)}]{Itzykson_b_1980}%
  \BibitemOpen
  \bibfield  {author} {\bibinfo {author} {\bibfnamefont {C.}~\bibnamefont
  {Itzykson}}\ and\ \bibinfo {author} {\bibfnamefont {J.-B.}\ \bibnamefont
  {Zuber}},\ }\href@noop {} {\emph {\bibinfo {title} {Quantum Field Theory}}}\
  (\bibinfo  {publisher} {McGraw-Hill Inc., New York},\ \bibinfo {year}
  {1980})\BibitemShut {NoStop}%
\bibitem [{\citenamefont {Dinu}\ \emph {et~al.}(2012)\citenamefont {Dinu},
  \citenamefont {Heinzl},\ and\ \citenamefont {Ilderton}}]{Dinu_2012}%
  \BibitemOpen
  \bibfield  {author} {\bibinfo {author} {\bibfnamefont {V.}~\bibnamefont
  {Dinu}}, \bibinfo {author} {\bibfnamefont {T.}~\bibnamefont {Heinzl}},\ and\
  \bibinfo {author} {\bibfnamefont {A.}~\bibnamefont {Ilderton}},\ }\href@noop
  {} {\bibfield  {journal} {\bibinfo  {journal} {Phys. Rev. D}\ }\textbf
  {\bibinfo {volume} {86}},\ \bibinfo {pages} {085037} (\bibinfo {year}
  {2012})}\BibitemShut {NoStop}%
\bibitem [{\citenamefont {Ilderton}\ and\ \citenamefont
  {Torgrimsson}(2013)}]{Ilderton_2013_b}%
  \BibitemOpen
  \bibfield  {author} {\bibinfo {author} {\bibfnamefont {A.}~\bibnamefont
  {Ilderton}}\ and\ \bibinfo {author} {\bibfnamefont {G.}~\bibnamefont
  {Torgrimsson}},\ }\href@noop {} {\bibfield  {journal} {\bibinfo  {journal}
  {Phys. Rev. D}\ }\textbf {\bibinfo {volume} {87}},\ \bibinfo {pages} {085040}
  (\bibinfo {year} {2013})}\BibitemShut {NoStop}%
\bibitem [{\citenamefont {King}(2021)}]{King_2021}%
  \BibitemOpen
  \bibfield  {author} {\bibinfo {author} {\bibfnamefont {B.}~\bibnamefont
  {King}},\ }\href@noop {} {\bibfield  {journal} {\bibinfo  {journal} {Phys.
  Rev. D}\ }\textbf {\bibinfo {volume} {103}},\ \bibinfo {pages} {036018}
  (\bibinfo {year} {2021})}\BibitemShut {NoStop}%
\bibitem [{\citenamefont {Edwards}\ and\ \citenamefont
  {Ilderton}(2021)}]{Edwards_2021}%
  \BibitemOpen
  \bibfield  {author} {\bibinfo {author} {\bibfnamefont {J.~P.}\ \bibnamefont
  {Edwards}}\ and\ \bibinfo {author} {\bibfnamefont {A.}~\bibnamefont
  {Ilderton}},\ }\href@noop {} {\bibfield  {journal} {\bibinfo  {journal}
  {Phys. Rev. D}\ }\textbf {\bibinfo {volume} {103}},\ \bibinfo {pages}
  {016004} (\bibinfo {year} {2021})}\BibitemShut {NoStop}%
\bibitem [{\citenamefont {Krachkov}(2024)}]{Krachkov_2024}%
  \BibitemOpen
  \bibfield  {author} {\bibinfo {author} {\bibfnamefont {P.~A.}\ \bibnamefont
  {Krachkov}},\ }\href@noop {} {\bibfield  {journal} {\bibinfo  {journal}
  {Phys. Rev. D}\ }\textbf {\bibinfo {volume} {109}},\ \bibinfo {pages}
  {076002} (\bibinfo {year} {2024})}\BibitemShut {NoStop}%
\bibitem [{\citenamefont {Di~Piazza}(2014)}]{Di_Piazza_2014}%
  \BibitemOpen
  \bibfield  {author} {\bibinfo {author} {\bibfnamefont {A.}~\bibnamefont
  {Di~Piazza}},\ }\href@noop {} {\bibfield  {journal} {\bibinfo  {journal}
  {Phys. Rev. Lett.}\ }\textbf {\bibinfo {volume} {113}},\ \bibinfo {pages}
  {040402} (\bibinfo {year} {2014})}\BibitemShut {NoStop}%
\bibitem [{\citenamefont {Di~Piazza}(2015)}]{Di_Piazza_2015}%
  \BibitemOpen
  \bibfield  {author} {\bibinfo {author} {\bibfnamefont {A.}~\bibnamefont
  {Di~Piazza}},\ }\href@noop {} {\bibfield  {journal} {\bibinfo  {journal}
  {Phys. Rev. A}\ }\textbf {\bibinfo {volume} {91}},\ \bibinfo {pages} {042118}
  (\bibinfo {year} {2015})}\BibitemShut {NoStop}%
\bibitem [{\citenamefont {Di~Piazza}(2021)}]{Di_Piazza_2021}%
  \BibitemOpen
  \bibfield  {author} {\bibinfo {author} {\bibfnamefont {A.}~\bibnamefont
  {Di~Piazza}},\ }\href@noop {} {\bibfield  {journal} {\bibinfo  {journal}
  {Phys. Rev. D}\ }\textbf {\bibinfo {volume} {103}},\ \bibinfo {pages}
  {076011} (\bibinfo {year} {2021})}\BibitemShut {NoStop}%
\bibitem [{\citenamefont {Mitter}(1975)}]{Mitter_1975}%
  \BibitemOpen
  \bibfield  {author} {\bibinfo {author} {\bibfnamefont {H.}~\bibnamefont
  {Mitter}},\ }\href@noop {} {\bibfield  {journal} {\bibinfo  {journal} {Acta
  Phys. Austriaca}\ }\textbf {\bibinfo {volume} {XIV}},\ \bibinfo {pages} {397}
  (\bibinfo {year} {1975})}\BibitemShut {NoStop}%
\bibitem [{\citenamefont {Ritus}(1985)}]{Ritus_1985}%
  \BibitemOpen
  \bibfield  {author} {\bibinfo {author} {\bibfnamefont {V.~I.}\ \bibnamefont
  {Ritus}},\ }\href@noop {} {\bibfield  {journal} {\bibinfo  {journal} {J. Sov.
  Laser Res.}\ }\textbf {\bibinfo {volume} {6}},\ \bibinfo {pages} {497}
  (\bibinfo {year} {1985})}\BibitemShut {NoStop}%
\bibitem [{\citenamefont {Ehlotzky}\ \emph {et~al.}(2009)\citenamefont
  {Ehlotzky}, \citenamefont {Krajewska},\ and\ \citenamefont
  {Kami\'{n}ski}}]{Ehlotzky_2009}%
  \BibitemOpen
  \bibfield  {author} {\bibinfo {author} {\bibfnamefont {F.}~\bibnamefont
  {Ehlotzky}}, \bibinfo {author} {\bibfnamefont {K.}~\bibnamefont
  {Krajewska}},\ and\ \bibinfo {author} {\bibfnamefont {J.~Z.}\ \bibnamefont
  {Kami\'{n}ski}},\ }\href@noop {} {\bibfield  {journal} {\bibinfo  {journal}
  {Rep. Prog. Phys.}\ }\textbf {\bibinfo {volume} {72}},\ \bibinfo {pages}
  {046401} (\bibinfo {year} {2009})}\BibitemShut {NoStop}%
\bibitem [{\citenamefont {Reiss}(2009)}]{Reiss_2009}%
  \BibitemOpen
  \bibfield  {author} {\bibinfo {author} {\bibfnamefont {H.~R.}\ \bibnamefont
  {Reiss}},\ }\href@noop {} {\bibfield  {journal} {\bibinfo  {journal} {Eur.
  Phys. J. D}\ }\textbf {\bibinfo {volume} {55}},\ \bibinfo {pages} {365}
  (\bibinfo {year} {2009})}\BibitemShut {NoStop}%
\bibitem [{\citenamefont {Torgrimsson}(2024)}]{Torgrimsson_2024}%
  \BibitemOpen
  \bibfield  {author} {\bibinfo {author} {\bibfnamefont {G.}~\bibnamefont
  {Torgrimsson}},\ }\href@noop {} {\bibfield  {journal} {\bibinfo  {journal}
  {Phys. Rev. D}\ }\textbf {\bibinfo {volume} {110}},\ \bibinfo {pages}
  {076012} (\bibinfo {year} {2024})}\BibitemShut {NoStop}%
\bibitem [{\citenamefont {Zhao}\ and\ \citenamefont {Tang}(2025)}]{Zhao_2025}%
  \BibitemOpen
  \bibfield  {author} {\bibinfo {author} {\bibfnamefont {Z.-d.}\ \bibnamefont
  {Zhao}}\ and\ \bibinfo {author} {\bibfnamefont {S.}~\bibnamefont {Tang}},\
  }\href@noop {} {\bibfield  {journal} {\bibinfo  {journal} {Phys. Rev. D}\
  }\textbf {\bibinfo {volume} {112}},\ \bibinfo {pages} {056033} (\bibinfo
  {year} {2025})}\BibitemShut {NoStop}%
\bibitem [{\citenamefont {Di~Piazza}\ and\ \citenamefont
  {Qu}(2026)}]{Di_Piazza_2026}%
  \BibitemOpen
  \bibfield  {author} {\bibinfo {author} {\bibfnamefont {A.}~\bibnamefont
  {Di~Piazza}}\ and\ \bibinfo {author} {\bibfnamefont {K.}~\bibnamefont {Qu}},\
  }\href@noop {} {\bibfield  {journal} {\bibinfo  {journal} {Phys. Rev. Lett.}\
  }\textbf {\bibinfo {volume} {136}},\ \bibinfo {pages} {085001} (\bibinfo
  {year} {2026})}\BibitemShut {NoStop}%
\bibitem [{\citenamefont {Di~Piazza}\ \emph {et~al.}(2018)\citenamefont
  {Di~Piazza}, \citenamefont {Tamburini}, \citenamefont {Meuren},\ and\
  \citenamefont {Keitel}}]{Di_Piazza_2018}%
  \BibitemOpen
  \bibfield  {author} {\bibinfo {author} {\bibfnamefont {A.}~\bibnamefont
  {Di~Piazza}}, \bibinfo {author} {\bibfnamefont {M.}~\bibnamefont
  {Tamburini}}, \bibinfo {author} {\bibfnamefont {S.}~\bibnamefont {Meuren}},\
  and\ \bibinfo {author} {\bibfnamefont {C.~H.}\ \bibnamefont {Keitel}},\
  }\href@noop {} {\bibfield  {journal} {\bibinfo  {journal} {Phys. Rev. A}\
  }\textbf {\bibinfo {volume} {98}},\ \bibinfo {pages} {012134} (\bibinfo
  {year} {2018})}\BibitemShut {NoStop}%
\bibitem [{Note1()}]{Note1}%
  \BibitemOpen
  \bibinfo {note} {It is important to point out that the assumption on the
  continuity of $\protect \bm {A}_{\perp }(\phi )$ is necessary even from a
  strictly mathematical point of view in order to carry out the change of
  variable from the time $t$ to the light-cone time $\phi $ in the
  integrals.}\BibitemShut {Stop}%
\bibitem [{\citenamefont {Jauch}\ and\ \citenamefont
  {Rohrlich}(1976)}]{Jauch_b_1976}%
  \BibitemOpen
  \bibfield  {author} {\bibinfo {author} {\bibfnamefont {J.~M.}\ \bibnamefont
  {Jauch}}\ and\ \bibinfo {author} {\bibfnamefont {F.}~\bibnamefont
  {Rohrlich}},\ }\href@noop {} {\emph {\bibinfo {title} {The Theory of Photons
  and Electrons}}}\ (\bibinfo  {publisher} {Springer, Berlin},\ \bibinfo {year}
  {1976})\BibitemShut {NoStop}%
\bibitem [{\citenamefont {Olver}\ \emph {et~al.}(2010)\citenamefont {Olver},
  \citenamefont {Lozier}, \citenamefont {Boisvert},\ and\ \citenamefont
  {Clark}}]{NIST_b_2010}%
  \BibitemOpen
  \bibinfo {editor} {\bibfnamefont {F.~W.~J.}\ \bibnamefont {Olver}}, \bibinfo
  {editor} {\bibfnamefont {D.~W.}\ \bibnamefont {Lozier}}, \bibinfo {editor}
  {\bibfnamefont {R.~F.}\ \bibnamefont {Boisvert}},\ and\ \bibinfo {editor}
  {\bibfnamefont {C.~W.}\ \bibnamefont {Clark}},\ eds.,\ \href@noop {} {\emph
  {\bibinfo {title} {NIST Handbook of Mathematical Functions}}}\ (\bibinfo
  {publisher} {Cambridge University Press, Cambridge, England},\ \bibinfo
  {year} {2010})\BibitemShut {NoStop}%
\bibitem [{\citenamefont {Baier}\ \emph {et~al.}(1998)\citenamefont {Baier},
  \citenamefont {Katkov},\ and\ \citenamefont {Strakhovenko}}]{Baier_b_1998}%
  \BibitemOpen
  \bibfield  {author} {\bibinfo {author} {\bibfnamefont {V.~N.}\ \bibnamefont
  {Baier}}, \bibinfo {author} {\bibfnamefont {V.~M.}\ \bibnamefont {Katkov}},\
  and\ \bibinfo {author} {\bibfnamefont {V.~M.}\ \bibnamefont {Strakhovenko}},\
  }\href@noop {} {\emph {\bibinfo {title} {Electromagnetic Processes at High
  Energies in Oriented Single Crystals}}}\ (\bibinfo  {publisher} {World
  Scientific, Singapore},\ \bibinfo {year} {1998})\BibitemShut {NoStop}%
\bibitem [{\citenamefont {Baier}\ and\ \citenamefont
  {Katkov}(2005)}]{Baier_2005}%
  \BibitemOpen
  \bibfield  {author} {\bibinfo {author} {\bibfnamefont {V.~N.}\ \bibnamefont
  {Baier}}\ and\ \bibinfo {author} {\bibfnamefont {V.~M.}\ \bibnamefont
  {Katkov}},\ }\href@noop {} {\bibfield  {journal} {\bibinfo  {journal} {Phys.
  Rep.}\ }\textbf {\bibinfo {volume} {409}},\ \bibinfo {pages} {261} (\bibinfo
  {year} {2005})}\BibitemShut {NoStop}%
\bibitem [{\citenamefont {Di~Piazza}\ and\ \citenamefont
  {Audagnotto}(2021)}]{Di_Piazza_2021_b}%
  \BibitemOpen
  \bibfield  {author} {\bibinfo {author} {\bibfnamefont {A.}~\bibnamefont
  {Di~Piazza}}\ and\ \bibinfo {author} {\bibfnamefont {G.}~\bibnamefont
  {Audagnotto}},\ }\href@noop {} {\bibfield  {journal} {\bibinfo  {journal}
  {Phys. Rev. D}\ }\textbf {\bibinfo {volume} {104}},\ \bibinfo {pages}
  {016007} (\bibinfo {year} {2021})}\BibitemShut {NoStop}%
\bibitem [{\citenamefont {Heinzl}\ \emph {et~al.}(2021)\citenamefont {Heinzl},
  \citenamefont {Ilderton},\ and\ \citenamefont {King}}]{Heinzl_2021}%
  \BibitemOpen
  \bibfield  {author} {\bibinfo {author} {\bibfnamefont {T.}~\bibnamefont
  {Heinzl}}, \bibinfo {author} {\bibfnamefont {A.}~\bibnamefont {Ilderton}},\
  and\ \bibinfo {author} {\bibfnamefont {B.}~\bibnamefont {King}},\ }\href@noop
  {} {\bibfield  {journal} {\bibinfo  {journal} {Phys. Rev. Lett.}\ }\textbf
  {\bibinfo {volume} {127}},\ \bibinfo {pages} {061601} (\bibinfo {year}
  {2021})}\BibitemShut {NoStop}%
\bibitem [{\citenamefont {Di~Piazza}\ \emph {et~al.}(2010)\citenamefont
  {Di~Piazza}, \citenamefont {Hatsagortsyan},\ and\ \citenamefont
  {Keitel}}]{Di_Piazza_2010}%
  \BibitemOpen
  \bibfield  {author} {\bibinfo {author} {\bibfnamefont {A.}~\bibnamefont
  {Di~Piazza}}, \bibinfo {author} {\bibfnamefont {K.~Z.}\ \bibnamefont
  {Hatsagortsyan}},\ and\ \bibinfo {author} {\bibfnamefont {C.~H.}\
  \bibnamefont {Keitel}},\ }\href@noop {} {\bibfield  {journal} {\bibinfo
  {journal} {Phys. Rev. Lett.}\ }\textbf {\bibinfo {volume} {105}},\ \bibinfo
  {pages} {220403} (\bibinfo {year} {2010})}\BibitemShut {NoStop}%
\bibitem [{\citenamefont {Torgrimsson}(2021)}]{Torgrimsson_2021}%
  \BibitemOpen
  \bibfield  {author} {\bibinfo {author} {\bibfnamefont {G.}~\bibnamefont
  {Torgrimsson}},\ }\href@noop {} {\bibfield  {journal} {\bibinfo  {journal}
  {Phys. Rev. Lett.}\ }\textbf {\bibinfo {volume} {127}},\ \bibinfo {pages}
  {111602} (\bibinfo {year} {2021})}\BibitemShut {NoStop}%
\bibitem [{\citenamefont {Elkina}\ \emph {et~al.}(2011)\citenamefont {Elkina},
  \citenamefont {Fedotov}, \citenamefont {{I. Yu. Kostyukov}}, \citenamefont
  {Legkov}, \citenamefont {Narozhny}, \citenamefont {Nerush},\ and\
  \citenamefont {Ruhl}}]{Elkina_2011}%
  \BibitemOpen
  \bibfield  {author} {\bibinfo {author} {\bibfnamefont {N.~V.}\ \bibnamefont
  {Elkina}}, \bibinfo {author} {\bibfnamefont {A.~M.}\ \bibnamefont {Fedotov}},
  \bibinfo {author} {\bibnamefont {{I. Yu. Kostyukov}}}, \bibinfo {author}
  {\bibfnamefont {M.~V.}\ \bibnamefont {Legkov}}, \bibinfo {author}
  {\bibfnamefont {N.~B.}\ \bibnamefont {Narozhny}}, \bibinfo {author}
  {\bibfnamefont {E.~N.}\ \bibnamefont {Nerush}},\ and\ \bibinfo {author}
  {\bibfnamefont {H.}~\bibnamefont {Ruhl}},\ }\href@noop {} {\bibfield
  {journal} {\bibinfo  {journal} {Phys. Rev. ST Accel. Beams}\ }\textbf
  {\bibinfo {volume} {14}},\ \bibinfo {pages} {054401} (\bibinfo {year}
  {2011})}\BibitemShut {NoStop}%
\bibitem [{\citenamefont {Neitz}\ and\ \citenamefont
  {Di~Piazza}(2013)}]{Neitz_2013}%
  \BibitemOpen
  \bibfield  {author} {\bibinfo {author} {\bibfnamefont {N.}~\bibnamefont
  {Neitz}}\ and\ \bibinfo {author} {\bibfnamefont {A.}~\bibnamefont
  {Di~Piazza}},\ }\href@noop {} {\bibfield  {journal} {\bibinfo  {journal}
  {Phys. Rev. Lett.}\ }\textbf {\bibinfo {volume} {111}},\ \bibinfo {pages}
  {054802} (\bibinfo {year} {2013})}\BibitemShut {NoStop}%
\bibitem [{\citenamefont {Di~Piazza}\ \emph {et~al.}(2008)\citenamefont
  {Di~Piazza}, \citenamefont {Hatsagortsyan},\ and\ \citenamefont
  {Keitel}}]{Di_Piazza_2008}%
  \BibitemOpen
  \bibfield  {author} {\bibinfo {author} {\bibfnamefont {A.}~\bibnamefont
  {Di~Piazza}}, \bibinfo {author} {\bibfnamefont {K.~Z.}\ \bibnamefont
  {Hatsagortsyan}},\ and\ \bibinfo {author} {\bibfnamefont {C.~H.}\
  \bibnamefont {Keitel}},\ }\href@noop {} {\bibfield  {journal} {\bibinfo
  {journal} {Phys. Rev. Lett.}\ }\textbf {\bibinfo {volume} {100}},\ \bibinfo
  {pages} {010403} (\bibinfo {year} {2008})}\BibitemShut {NoStop}%
\bibitem [{\citenamefont {Di~Piazza}(2008)}]{Di_Piazza_2008_a}%
  \BibitemOpen
  \bibfield  {author} {\bibinfo {author} {\bibfnamefont {A.}~\bibnamefont
  {Di~Piazza}},\ }\href@noop {} {\bibfield  {journal} {\bibinfo  {journal}
  {Lett. Math. Phys.}\ }\textbf {\bibinfo {volume} {83}},\ \bibinfo {pages}
  {305} (\bibinfo {year} {2008})}\BibitemShut {NoStop}%
\bibitem [{\citenamefont {Wistisen}(2014)}]{Wistisen_2014}%
  \BibitemOpen
  \bibfield  {author} {\bibinfo {author} {\bibfnamefont {T.~N.}\ \bibnamefont
  {Wistisen}},\ }\href@noop {} {\bibfield  {journal} {\bibinfo  {journal}
  {Phys. Rev. D}\ }\textbf {\bibinfo {volume} {90}},\ \bibinfo {pages} {125008}
  (\bibinfo {year} {2014})}\BibitemShut {NoStop}%
\bibitem [{\citenamefont {Karsch}\ \emph {et~al.}(2008)\citenamefont {Karsch},
  \citenamefont {Major}, \citenamefont {F\"{u}l\"{o}p}, \citenamefont {Ahmad},
  \citenamefont {Wang}, \citenamefont {Henig}, \citenamefont {Kruber},
  \citenamefont {Weingartner}, \citenamefont {Siebold}, \citenamefont {Hein},
  \citenamefont {Wandt}, \citenamefont {Klingebiel}, \citenamefont {Osterhoff},
  \citenamefont {Horlein},\ and\ \citenamefont {Krausz}}]{Karsch_2008}%
  \BibitemOpen
  \bibfield  {author} {\bibinfo {author} {\bibfnamefont {S.}~\bibnamefont
  {Karsch}}, \bibinfo {author} {\bibfnamefont {Z.}~\bibnamefont {Major}},
  \bibinfo {author} {\bibfnamefont {J.}~\bibnamefont {F\"{u}l\"{o}p}}, \bibinfo
  {author} {\bibfnamefont {I.}~\bibnamefont {Ahmad}}, \bibinfo {author}
  {\bibfnamefont {T.-J.}\ \bibnamefont {Wang}}, \bibinfo {author}
  {\bibfnamefont {A.}~\bibnamefont {Henig}}, \bibinfo {author} {\bibfnamefont
  {S.}~\bibnamefont {Kruber}}, \bibinfo {author} {\bibfnamefont
  {R.}~\bibnamefont {Weingartner}}, \bibinfo {author} {\bibfnamefont
  {M.}~\bibnamefont {Siebold}}, \bibinfo {author} {\bibfnamefont
  {J.}~\bibnamefont {Hein}}, \bibinfo {author} {\bibfnamefont {C.}~\bibnamefont
  {Wandt}}, \bibinfo {author} {\bibfnamefont {S.}~\bibnamefont {Klingebiel}},
  \bibinfo {author} {\bibfnamefont {J.}~\bibnamefont {Osterhoff}}, \bibinfo
  {author} {\bibfnamefont {R.}~\bibnamefont {Horlein}},\ and\ \bibinfo {author}
  {\bibfnamefont {F.}~\bibnamefont {Krausz}},\ }in\ \href@noop {} {\emph
  {\bibinfo {booktitle} {Advanced Solid-State Photonics}}}\ (\bibinfo
  {publisher} {Optica Publishing Group},\ \bibinfo {year} {2008})\ p.\ \bibinfo
  {pages} {WF1}\BibitemShut {NoStop}%
\bibitem [{\citenamefont {Ter-Mikaelian}(1972)}]{Ter-Mikaelian_b_1972}%
  \BibitemOpen
  \bibfield  {author} {\bibinfo {author} {\bibfnamefont {M.~L.}\ \bibnamefont
  {Ter-Mikaelian}},\ }\href@noop {} {\emph {\bibinfo {title} {High-Energy
  Electromagnetic Processes in Condensed Matter}}}\ (\bibinfo  {publisher}
  {Wiley-Interscience, Toronto},\ \bibinfo {year} {1972})\BibitemShut {NoStop}%
\end{thebibliography}

\end{document}